\documentclass[twocolumn,superscriptaddress,longbibliography]{revtex4-2}
\usepackage{amsmath,amssymb,amsfonts}
\usepackage{xcolor,graphicx}
\newcommand{\rev}[1]{#1}

\usepackage[bookmarks=false,linkcolor=cyan,urlcolor=blue,colorlinks,citecolor=cyan]{hyperref}

\newcommand{\ket}[1]{|#1\rangle}

\begin{document}

\title{Multimode Purcell Filter for Superconducting-Qubit Reset and Readout with Intrinsic Purcell Protection}

\author{Xu-Yang Gu}
\thanks{These authors contributed equally to this work.}
\affiliation{Beijing National Laboratory for Condensed Matter Physics, Institute of Physics, Chinese Academy of Sciences, Beijing 100190, China}
\affiliation{School of Physical Sciences, University of Chinese Academy of Sciences, Beijing 100049, China}

\author{Da'er Feng}
\thanks{These authors contributed equally to this work.}
\affiliation{Beijing National Laboratory for Condensed Matter Physics, Institute of Physics, Chinese Academy of Sciences, Beijing 100190, China}
\affiliation{School of Physical Sciences, University of Chinese Academy of Sciences, Beijing 100049, China}

\author{Zhen-Yu Peng}
\affiliation{Beijing National Laboratory for Condensed Matter Physics, Institute of Physics, Chinese Academy of Sciences, Beijing 100190, China}
\affiliation{School of Physical Sciences, University of Chinese Academy of Sciences, Beijing 100049, China}

\author{Gui-Han Liang}
\affiliation{Beijing National Laboratory for Condensed Matter Physics, Institute of Physics, Chinese Academy of Sciences, Beijing 100190, China}
\affiliation{School of Physical Sciences, University of Chinese Academy of Sciences, Beijing 100049, China}

\author{Yang He}
\affiliation{Beijing National Laboratory for Condensed Matter Physics, Institute of Physics, Chinese Academy of Sciences, Beijing 100190, China}
\affiliation{School of Physical Sciences, University of Chinese Academy of Sciences, Beijing 100049, China}

\author{Yongxi Xiao}
\affiliation{Beijing National Laboratory for Condensed Matter Physics, Institute of Physics, Chinese Academy of Sciences, Beijing 100190, China}
\affiliation{School of Physical Sciences, University of Chinese Academy of Sciences, Beijing 100049, China}

\author{Ming-Chuan Wang}
\affiliation{Beijing National Laboratory for Condensed Matter Physics, Institute of Physics, Chinese Academy of Sciences, Beijing 100190, China}
\affiliation{School of Physical Sciences, University of Chinese Academy of Sciences, Beijing 100049, China}

\author{Yu Yan}
\affiliation{Beijing National Laboratory for Condensed Matter Physics, Institute of Physics, Chinese Academy of Sciences, Beijing 100190, China}
\affiliation{School of Physics, Northwest University, Xi’an 710127, China}

\author{Bing-Jie Chen}
\affiliation{Beijing National Laboratory for Condensed Matter Physics, Institute of Physics, Chinese Academy of Sciences, Beijing 100190, China}
\affiliation{School of Physical Sciences, University of Chinese Academy of Sciences, Beijing 100049, China}

\author{Zheng-Yang Mei}
\affiliation{Beijing National Laboratory for Condensed Matter Physics, Institute of Physics, Chinese Academy of Sciences, Beijing 100190, China}
\affiliation{School of Physical Sciences, University of Chinese Academy of Sciences, Beijing 100049, China}

\author{Yi-Zhou Bu}
\affiliation{Beijing National Laboratory for Condensed Matter Physics, Institute of Physics, Chinese Academy of Sciences, Beijing 100190, China}
\affiliation{School of Physical Sciences, University of Chinese Academy of Sciences, Beijing 100049, China}

\author{Jia-Chi Zhang}
\affiliation{Beijing National Laboratory for Condensed Matter Physics, Institute of Physics, Chinese Academy of Sciences, Beijing 100190, China}
\affiliation{School of Physical Sciences, University of Chinese Academy of Sciences, Beijing 100049, China}

\author{Jia-Cheng Song}
\affiliation{Beijing National Laboratory for Condensed Matter Physics, Institute of Physics, Chinese Academy of Sciences, Beijing 100190, China}
\affiliation{School of Physical Sciences, University of Chinese Academy of Sciences, Beijing 100049, China}

\author{Cheng-Lin Deng}
\affiliation{Beijing National Laboratory for Condensed Matter Physics, Institute of Physics, Chinese Academy of Sciences, Beijing 100190, China}
\affiliation{School of Physical Sciences, University of Chinese Academy of Sciences, Beijing 100049, China}

\author{Yun-Hao Shi}
\affiliation{Beijing National Laboratory for Condensed Matter Physics, Institute of Physics, Chinese Academy of Sciences, Beijing 100190, China}

\author{Xiaohui Song}
\affiliation{Beijing National Laboratory for Condensed Matter Physics, Institute of Physics, Chinese Academy of Sciences, Beijing 100190, China}

\author{Dongning Zheng}
\affiliation{Beijing National Laboratory for Condensed Matter Physics, Institute of Physics, Chinese Academy of Sciences, Beijing 100190, China}
\affiliation{School of Physical Sciences, University of Chinese Academy of Sciences, Beijing 100049, China}
\affiliation{Hefei National Laboratory, Hefei 230088, China}
\affiliation{Songshan Lake Materials Laboratory, Dongguan, Guangdong 523808, China}
	
\author{Kai Xu}
\email{kaixu@iphy.ac.cn}
\affiliation{Beijing National Laboratory for Condensed Matter Physics, Institute of Physics, Chinese Academy of Sciences, Beijing 100190, China}
\affiliation{School of Physical Sciences, University of Chinese Academy of Sciences, Beijing 100049, China}
\affiliation{Beijing Key Laboratory of Fault-Tolerant Quantum Computing, Beijing Academy of Quantum Information Sciences, Beijing 100193, China}
\affiliation{Hefei National Laboratory, Hefei 230088, China}
\affiliation{Songshan Lake Materials Laboratory, Dongguan, Guangdong 523808, China}

\author{Zhongcheng Xiang}
\email{zcxiang@iphy.ac.cn}
\affiliation{Beijing National Laboratory for Condensed Matter Physics, Institute of Physics, Chinese Academy of Sciences, Beijing 100190, China}
\affiliation{School of Physical Sciences, University of Chinese Academy of Sciences, Beijing 100049, China}
\affiliation{Hefei National Laboratory, Hefei 230088, China}

\author{Heng Fan}
\email{hfan@iphy.ac.cn}
\affiliation{Beijing National Laboratory for Condensed Matter Physics, Institute of Physics, Chinese Academy of Sciences, Beijing 100190, China}
\affiliation{School of Physical Sciences, University of Chinese Academy of Sciences, Beijing 100049, China}
\affiliation{Beijing Key Laboratory of Fault-Tolerant Quantum Computing, Beijing Academy of Quantum Information Sciences, Beijing 100193, China}
\affiliation{Hefei National Laboratory, Hefei 230088, China}
\affiliation{Songshan Lake Materials Laboratory, Dongguan, Guangdong 523808, China}


\begin{abstract}
Efficient qubit reset and leakage reduction are essential for scalable superconducting quantum computing, particularly in the context of quantum error correction. However, such operations often require additional on-chip components. Here, we propose and experimentally demonstrate a hardware-efficient approach to qubit reset and readout using a multi-mode Purcell filter in a superconducting quantum circuit.
We exploit the inherent multi-mode structure of a coplanar waveguide resonator, using its fundamental and second-order modes for qubit reset and readout, respectively, thereby avoiding additional components. 
Implemented in a flip-chip architecture, our device achieves unconditional reset with residual excitation below 1\% in 220 ns, and a leakage reduction unit that selectively resets the second excited state within 62 ns with a residual $|f\rangle$ population of 6.1\%, accounting for the readout error. 
\rev{Despite the qubits being directly coupled to the filter in our configuration, the measured relaxation times are not degraded owing to intrinsic Purcell protection provided by an auxiliary mode.}
To our knowledge, this is the first experimental trial that exploits different-order modes of a microwave resonator for distinct qubit operations, representing a new direction toward scalable, hardware-efficient quantum processor design.
\end{abstract}
\maketitle

\section{Introduction}
\label{sec:intro}

Fault-tolerant superconducting quantum computing requires sophisticated manipulation of up to a million physical qubits \cite{fowlerSurfaceCodesPractical2012,gidneyHowFactor20482025}.
To advance toward this goal, two complementary directions have attracted considerable attention.
One involves the development of distributed quantum network architectures \cite{axlineOndemandQuantumState2018,campagne-ibarcqDeterministicRemoteEntanglement2018,kurpiersDeterministicQuantumState2018,leungDeterministicBidirectionalCommunication2019,magnardMicrowaveQuantumLink2020,burkhartErrorDetectedStateTransfer2021,zhongDeterministicMultiqubitEntanglement2021a,niuLowlossInterconnectsModular2023a,storzLoopholefreeBellInequality2023,qiuDeterministicQuantumState2025,qiuThermalnoiseresilientMicrowaveQuantum2025}, while the other focuses on maximizing the number of physical qubits integrated on a single chip.
Achieving the latter requires minimizing the number of on-chip components while maintaining sufficient controllability of the qubits.
However, enhanced controllability often requires additional physical components on the chip.
For example, operations such as fast unconditional reset \cite{reedFastResetSuppressing2010a,magnardFastUnconditionalAllMicrowave2018,mcewenRemovingLeakageinducedCorrelated2021,zhouRapidUnconditionalParametric2021,kimFastUnconditionalReset2024,dingMultipurposeArchitectureFast2025} and leakage reduction unit (LRU) \cite{aliferisFaultTolerantQuantumComputation2006,battistelHardwareEfficientLeakageReductionScheme2021,mcewenRemovingLeakageinducedCorrelated2021,yangCouplerAssistedLeakageReduction2024a,kimFastUnconditionalReset2024,thorbeckHighfidelityGatesTransmon2024}, which are essential for quantum error correction \cite{fowlerSurfaceCodesPractical2012,reedRealizationThreequbitQuantum2012,googlequantumaiExponentialSuppressionBit2021a,krinnerRealizingRepeatedQuantum2022,googlequantumaiSuppressingQuantumErrors2023,googlequantumaiandcollaboratorsQuantumErrorCorrection2025}, typically requires a dedicated dissipative channel whose frequency lies below the qubit frequency \cite{dingMultipurposeArchitectureFast2025,kimFastUnconditionalReset2024,thorbeckHighfidelityGatesTransmon2024}.
Despite the existence of several alternative approaches, such as repurposing the readout resonator as a reset channel by placing its frequency below the qubit band \cite{reedFastResetSuppressing2010a,mcewenRemovingLeakageinducedCorrelated2021,miaoOvercomingLeakageQuantum2023}, employing parametrically activated interactions \cite{zhouRapidUnconditionalParametric2021,lacroixFastFluxActivatedLeakage2025}, or utilizing cavity-assisted Raman transitions \cite{magnardFastUnconditionalAllMicrowave2018,battistelHardwareEfficientLeakageReductionScheme2021,marquesAllMicrowaveLeakageReduction2023}, each comes with its own limitations.
The latter two typically require complex calibration to avoid spurious transitions and are sensitive to noise and parameter drift, while the former may introduce undesired effects such as measurement-induced mixing \cite{dumasMeasurementInducedTransmonIonization2024}.

To realize qubit reset and leakage reduction operations without introducing additional on-chip components, we propose assigning multiple functions to a single physical element. 
For example, a half-wavelength coplanar waveguide resonator is inherently a multi-mode system, supporting a fundamental mode at frequency $f_0$ and higher-order harmonics at $f_m=(m+1)f_0$. 
Harnessing these different-order modes for distinct qubit operations offers a promising route toward improving circuit scalability. 
In this work, we design and experimentally demonstrate a multi-mode Purcell filter, where the fundamental mode is used for qubit reset and the second-order mode for readout. 
This approach offers a new route toward hardware-efficient circuit design and scalable superconducting quantum architectures.

\begin{figure*}[ht!]
\centering
\includegraphics{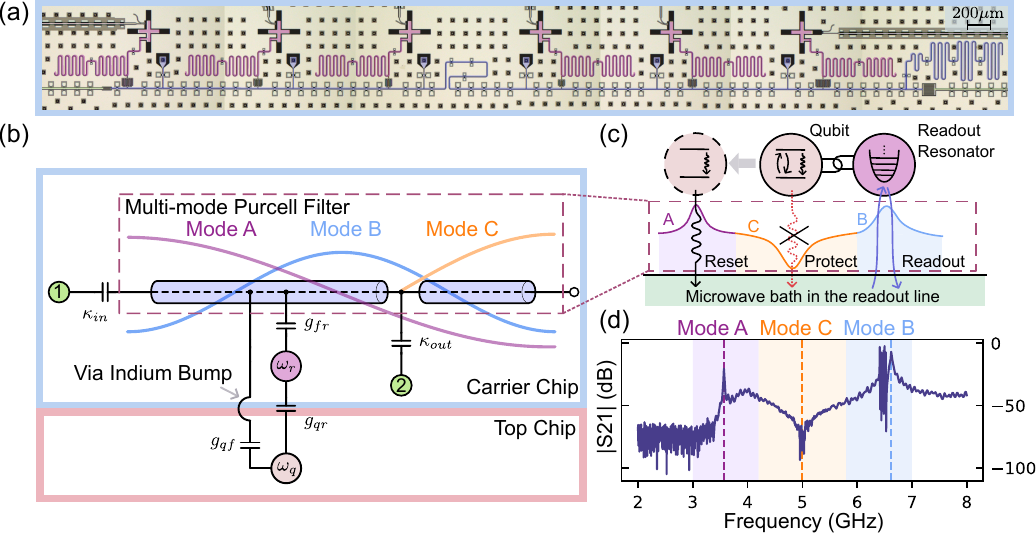}
\caption{
Overview of the multi-mode Purcell filter.
(a) False-colored photograph of the carrier chip of the device, including the multi-mode Purcell filter (blue) shared by six readout resonators (pink), and the input and output feedlines (light green).
The full device is a flip-chip superconducting quantum circuit consisting of a top qubit chip and a bottom carrier chip.
(b) Circuit model of the multi-mode filter. The filter is an open-circuited transmission-line resonator, where mode A is its fundamental mode and mode B is its second-order mode. 
Mode C is a $\lambda/4$ transmission-line stub that introduces a notch near 5 GHz. 
(c) Schematic illustration of the working principle of the multi-mode Purcell filter. The filter modifies the microwave bath seen by the qubit. 
Modes A and B serve as channels for qubit reset and readout, respectively, while mode C provides a notch filtering around qubit operation frequency, suppressing energy leakage.
(d) Transmission spectrum $S_{21}$ of the readout chain. The peak around 3.6 GHz corresponds to mode A, and the passband around 6.6 GHz corresponds to mode B. 
The dip around 5 GHz is the stopband produced by the quarter-wavelength stub (mode C). The readout resonators manifest themselves as dips within the passband constructed by mode B. 
The data are obtained using a vector network analyzer.
}
\label{fig_1}
\end{figure*}

The proposed scheme is implemented by coupling the qubit and its readout resonator to a multi-mode Purcell filter, which supports a fundamental mode at approximately 3.6 GHz and a higher-order mode at around 6.6 GHz, serving as channels for qubit reset and readout, respectively. 
In addition, an auxiliary mode is carefully embedded in our filter to provide intrinsic Purcell protection \cite{sunadaFastReadoutReset2022,springFastMultiplexedSuperconductingQubit2025,yenInterferometricPurcellSuppression2025} for the qubit around its operation frequency. 
We experimentally demonstrate the performance of this multi-mode Purcell filter in a flip-chip superconducting circuit, in which we demonstrate a fast unconditional reset with residual excitation less than 1\% within 220 ns, and a leakage reduction unit that selectively resets the transmon’s second excited state within 62 ns. 
\rev{The Purcell-limited relaxation time $T_p$ of the qubit is also investigated numerically, with the simulations predicting $T_p$ exceeding 500~$\mu$s over a bandwidth of approximately 700~MHz, provided that the internal quality factors of the readout and filter resonators reach the level of $10^6$.}

\section{ENGINEERING A MULTI-MODE PURCELL FILTER}
\label{sec:concept}

The overview of our multi-mode Purcell filter is illustrated in Fig.~\ref{fig_1}.
Leveraging the multi-mode feature of the open-ended transmission-line resonator, three carefully engineered modes are embedded in the filter. These modes, labeled as A, B, and C in Fig.~\ref{fig_1}, are responsible for the qubit reset, readout, and additional Purcell protection, respectively. 
Mode A is the fundamental mode of this open-ended transmission-line resonator, with a resonance frequency of $\omega_A/2\pi$ = 3.6 GHz, which lies below the maximum qubit frequency.
As a result, we can reset the qubit by tuning down its frequency to that of mode A, by which the qubit energy is first transferred to mode A and then rapidly dissipates due to the large linewidth of the filter. 
Mode B is the second-order mode of the filter and has a resonance frequency of $\omega_B/2\pi$ = 6.6 GHz, which is aligned with the readout frequency. 
Taking advantage of this mode, readout pulse injected through the input feedline can interact with the readout resonator and gain information about the qubit state.
Finally, by moving the output capacitor to the position $\lambda_q/4$ away from the open end of the filter, a quarter-wavelength stub (mode C) is formed.
This stub creates a stopband near the qubit frequency (see Fig.~\ref{fig_1}(d)), thereby suppressing Purcell decay through the filter.
In other words, mode C in this distributed filter resonator corresponds to the electromagnetic mode that shows a voltage node at the position of the output capacitor. 
In this way, mode C can be decoupled from the microwave bath in the readout line. 
If the frequency of mode C is designed to be around that of the qubit, the energy leakage of the qubit through the output capacitor can be suppressed, and the Purcell-limited relaxation time of the qubit can be greatly enhanced.

Together, these three modes define a specific electromagnetic environment that enables qubit reset, readout, and intrinsic Purcell protection, as can be seen from the transmission spectrum in Fig.~\ref{fig_1}(d) and the calculated Purcell-limited relaxation time in Fig.~\ref{fig_pfpro}.

The experimental realization of the aforementioned multi-mode Purcell filter is carried out on a flip-chip superconducting circuit. As shown in Fig.~\ref{fig_1}(b), the qubits are fabricated on the top chip, while the readout resonators and the multimode Purcell filter are fabricated on the carrier chip.
Here, the multi-mode filter is shared among six qubits and their corresponding readout resonators, and their parameters are listed in Table~\ref{tab:params} in Appendix~\ref{sec_device}.
To characterize the performance of the multimode Purcell filter, the transmission spectrum $S_{21}$ between the input and output ports of the readout chain is measured using a vector network analyzer, and the result is plotted in Fig.~\ref{fig_1}(d). 
The peak around 3.6 GHz is the first-order mode (i.e., mode A) of this open-end transmission-line resonator filter, and another peak around 6.6 GHz corresponds to the second-order mode (i.e., mode B) of the filter. 
The dips located around the peak of mode B are manifestations of the readout resonators whose frequencies are placed within the passband of the filter. 
Here, the frequency of the second-order mode (mode B) is not exactly twice that of the fundamental mode (mode A). This deviation arises because the branches connecting the $C_{qf}$ act as open-circuited stubs within the filter, producing different frequency shifts for the fundamental and higher-order modes. A detailed quantitative analysis of this effect is provided in Appendix~\ref{sec:f_shift}.
While modes A and B perform themselves as peaks in the transmission spectrum, mode C contributes to a stopband around 5 GHz in the transmission spectrum, which prevents energy around this frequency from radiating to the readout line. 
In this way, additional Purcell protection can be applied to the qubit when it is tuned into the stopband built by mode C. 
As shown in Fig.~\ref{fig_1}(d), the S21 dip associated with mode C reaches the measurement floor of our vector network analyzer, making it difficult to determine the exact suppression level quantitatively. 
Nevertheless, from the available data we can conservatively estimate that mode C provides at least 20 dB of suppression.
By fitting the S21 data with the equation derived from input-output theory, the parameters of these modes can be extracted, and the results are listed in Table~\ref{table1} (see Appendix~\ref{sec_filterFit} for details).

Note that here mode A serves as a low-frequency mode to which the qubits couple.
In this sense, it may lead to measurement-induced ionization \cite{dumasMeasurementInducedTransmonIonization2024} of the qubit during the readout process.
However, since the frequency of mode A ($\sim$ 3.6GHz) is far detuned from the readout pulse ($\sim$ 6.5GHz), the photon number of mode A is much lower than the threshold to induce measurement-induced ionization. Consequently, the contribution of this mode to measurement-induced ionization is expected to be negligible.

\begin{table}[b]
\caption{\label{table1}%
The frequency and linewidth of modes A, B obtained from the fit to the transmission spectrum shown in Fig.~\ref{fig_1}(d). Mode A and mode B are fitted individually, and the fit function is obtained using input-output formalism. See Appendix~\ref{sec_filterFit} for more details.
}
\begin{ruledtabular}
\begin{tabular}{cccc}
\textrm{$\omega_A/2\pi$ (GHz)}&
\textrm{$\kappa_A/2\pi$ (MHz)}&
\multicolumn{1}{c}{$\omega_B/2\pi$ (GHz)}&
\textrm{$\kappa_B/2\pi$ (MHz)}\\
\colrule
3.567 & 5.4 & 6.583 & 20.2\\
\end{tabular}
\end{ruledtabular}
\end{table}

\section{Unconditional reset using the fundamental mode of the filter}
\label{sec:reset}

An essential feature of our multi-mode Purcell filter is that its fundamental mode (mode A), centered around 3.6 GHz, can be utilized to implement unconditional qubit reset. 
As shown in Fig.~\ref{fig_1}(b), the qubit is coupled to mode A through both a direct capacitive path and an indirect path mediated by the readout resonator. 
However, the indirect coupling is negligible due to the large detuning, approximately 3 GHz, between mode A and the readout resonator. Consequently, direct coupling is crucial for enabling efficient energy transfer from the qubit to the filter mode, which enables the implementation of the reset protocol.
Since the frequency of mode A is lower than the qubit transition frequency $\omega_{eg}/2\pi$, the qubit can be tuned into resonance with mode A via direct flux control. 
In this way, the energy of the qubit can be transfered to the filter mode, and then rapidly dissipates due to the relatively large linewidth of the filter. 
In addition to resetting the first excited state $|e\rangle$ into the ground state $|g\rangle$, this method can be extended to reset both the $|e\rangle$ and the $|f\rangle$ of the qubit by cascading the f-e and e-g reset pulses, which bring the f-e and e-g transitions into resonance with mode A successively.

\begin{figure*}[ht!]
\centering
\includegraphics{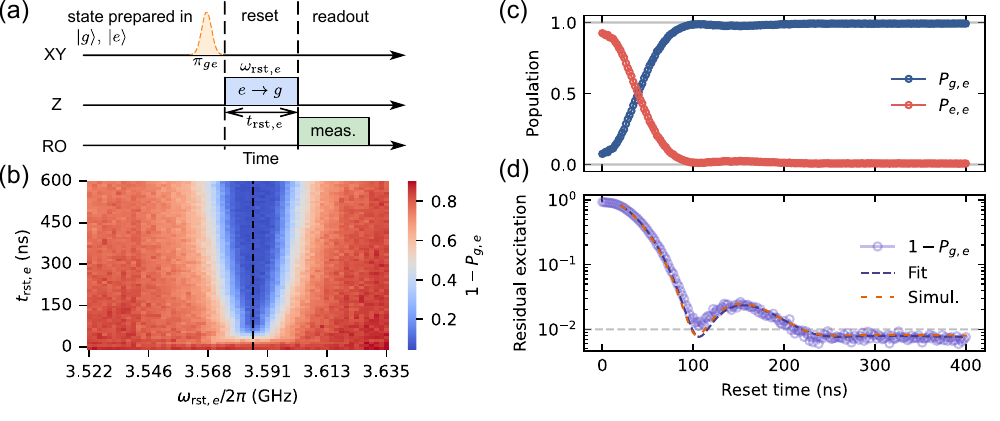}
\vspace*{-5mm}
\caption{
Fast unconditional reset via the fundamental mode of the filter. 
(a) Schematic of the pulse sequence used to characterize the reset protocol. 
(b) Reset error $1-P_{g,e}$ versus the reset frequency $\omega _{\mathrm{rst},e}/2\pi$ and the reset time $t _{\mathrm{rst},e}$. The optimal reset frequency $\omega _{\mathrm{rst},e}/2\pi$ is found to be 3.585 GHz (dark dashed line).
(c)(d) The qubit population $P_{g(e)}$ and the residual excitation $1-P_{g,e}$ versus the reset time $t _{\mathrm{rst},e}$. The qubit is first prepared in $|e\rangle$, and the reset frequency $\omega _{\mathrm{rst},e}/2\pi$ is chosen to be 3.585 GHz (dark dashed line in Fig.~\ref{fig_2}(b)). 15000 shots were taken per point. The purple dashed line is the fit using Eq.~(\ref{eq:reset}), and the orange dashed line is the result of the numerical simulation based on Eq.~(\ref{eq:master_equ}).
The horizontal dashed line corresponds to $1-P_{g,e}$ = 1\%, and the residual excitation $1-P_{g,e}$ is all below 1\% for reset time $t _{\mathrm{rst},e}>220$ ns.
}
\label{fig_2}
\end{figure*}

We use the pulse sequence shown in Fig.~\ref{fig_2}(a) to characterize the above reset protocol on Q4 of the device (see Appendix~\ref{sec_readout} for its readout characterization, with $F_{0}=98.9\%$ and $F_{1}=91.1\%$). 
Further analysis of the readout performance is presented in Appendix~\ref{sec_readout}.
The qubit is optionally excited to $|e\rangle$ immediately before applying the reset pulse, and during the reset pulse the qubit is rapidly tuned to the reset frequency $\omega_{\mathrm{rst},e}$ for the duration of $t _{\mathrm{rst},e}$ and then tuned back. 
After that, the final state of the qubit is extracted by dispersive measurement. 
We scan the reset frequency $\omega_{\mathrm{rst},e}$ and the reset time $t _{\mathrm{rst},e}$, and measure the resulting qubit state. The result is plotted in Fig.~\ref{fig_2}(b). 
We can see that the optimal reset frequency is around 3.585 GHz (dark dashed line in Fig.~\ref{fig_2}(b)), which coincides with the first transmission peak of the S21 spectrum shown in Fig.~\ref{fig_1}(d) and Table~\ref{table1}.

We adopt the reset frequency $\omega_{\mathrm{rst},e}/2\pi$ = 3.585 GHz obtained from Fig.~\ref{fig_2}(b), calibrate the reset pulse shape, and then characterize the final state population $P_{g(e),e}$ ($P_{i,j}$  is defined as the $|i\rangle$ population of the transmon initialized in $|j\rangle$) as a function of the reset time $t_{\mathrm{rst},e}$ (see Fig.~\ref{fig_2}(c)). The residual excitation $1-P_{g,e}$ is also computed and shown in Fig.~\ref{fig_2}(d). The result shows that the residual excitation can be reduced to less than 1\% within 220 ns. Here, the $1-P_{g,e}$ versus time can be described by \cite{zhouRapidUnconditionalParametric2021}
\begin{equation}
1-P_{g,e}=\begin{cases}
	\mathrm{if}\ \left| g_{qf} \right|<\kappa _f/4:\\
	e^{-\kappa _ft/2}\left[ \cosh \left( M_1t \right) +\frac{\kappa _f}{4M_1}\sinh \left( M_1t \right) \right] ^2,\\
	\mathrm{if}\ \left| g_{qf} \right|=\kappa _f/4:\\
	e^{-\kappa _ft/2}\left( \kappa _ft/4+1 \right) ^2,\\
	\mathrm{if}\ \left| g_{qf} \right|>\kappa _f/4:\\
	e^{-\kappa _ft/2}\left[ \cos \left( M_2t \right) +\frac{\kappa _f}{4M_2}\sin \left( M_2t \right) \right] ^2,\\
\end{cases}
\label{eq:reset}
\end{equation}
where $\kappa_f/2\pi$ is the linewidth of the filter mode, $g_{qf}/2\pi$ is the coupling strength between the qubit and the filter, $M_1=\sqrt{\kappa _{f}^{2}-16\left| g_{qf} \right|^2}/4$ and $M_2=\sqrt{16\left| g_{qf} \right|^2-\kappa _{f}^{2}}/4$. 
Depending on the filter linewidth $\kappa_f/2\pi$ and the coupling strength $g_{qf}/2\pi$, three different regimes are possible, corresponding to overdamped, critically damped, and underdamped oscillations. 
The result shown in Fig.~\ref{fig_2}(d) corresponds to the underdamped regime where $g_{qf}>\kappa _f/4$. 
From the fit of the data using Eq.~(\ref{eq:reset}), we extract the filter (mode A) linewidth $\kappa_f/2\pi$ = 8.5 MHz and the coupling strength $g_{qf}/2\pi$ = 3.9 MHz.  
From the fit, we also extract the steady-state residual excitation $P_{\mathrm{exc}}^{\mathrm{s}.\mathrm{s}}$ = 0.8\%, which is mainly caused by the readout error $1-F_0$=1.1\%.

\begin{figure*}[ht!]
\centering
\includegraphics{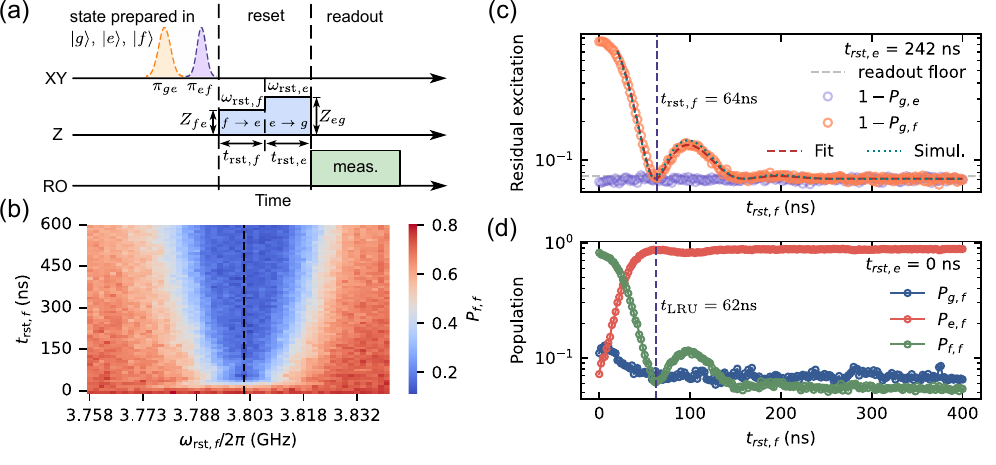}
\vspace*{-5mm}
\caption{
Demonstration of the f-e-g cascaded reset scheme. 
(a) Pulse sequence used to reset both the first and second excited states of a transmon qubit. 
(b) The landscape of $P_{f,f}$ versus the f-e reset time $t_{\mathrm{rst},f}$ and the reset frequency $\omega _{\mathrm{rst},f}/2\pi$. The optimal $\omega _{\mathrm{rst},f}/2\pi$ used to reset the $|f\rangle$ is found to be 3.801 GHz (dark dashed line).
(c)	The residual excitation $1-P_{g,f\left( e \right)}
$ versus the reset time $t_{\mathrm{rst},f}$, with $\omega _{\mathrm{rst},f}/2\pi$ = 3.801 GHz, $\omega _{\mathrm{rst},e}/2\pi$ = 3.585 GHz, and $t_{\mathrm{rst},e}$ = 242 ns. $1-P_{g,f}$ reaches its first minimum at $t_{\mathrm{rst},f}$ = 64 ns (the vertical purple dashed line). The red dashed line is the fit to Eq.~(\ref{eq:reset}), and the green dotted line is the result of the numerical simulation based on Eq.~(\ref{eq:master_equ}). 
The readout floor $1-F_0$ is indicated by the horizontal dashed line. 
(d) The resulting population of a transmon initialized in $|f\rangle$ after an LRU operation, where $t_{\mathrm{rst},e}
$ = 0, $\omega _{\mathrm{rst},f}/2\pi 
$ = 3.801 GHz, and $t_{\mathrm{rst},f}
$ is swept. The $P_{f,f}$ reaches its first minimum at $t_{\mathrm{rst},f}$ = 62 ns (the vertical purple dashed line).
}
\label{fig_3}
\end{figure*}

Next, we demonstrate that the above reset scheme can be extended to reset both the $|e\rangle$ and the $|f\rangle$ of the qubit by cascading the f-e and e-g reset pulses. To extend the scheme to reset the second-excited state $|f\rangle$, we first apply a Z pulse with amplitude $Z_{fe}$ to bring $\omega _{fe}/2\pi$ into resonance with mode A for a duration of $t_{\mathrm{rst},f}$, and then apply a second Z pulse with amplitude $Z_{eg}$ to bring $\omega _{eg}/2\pi$ into resonance with mode A for a duration of $t_{\mathrm{rst},e}$. 
The qubit frequencies during the f-e and e-g reset processes are labeled as $\omega _{\mathrm{rst},f}$ and $\omega _{\mathrm{rst},e}$, respectively.
We benchmark the efficiency of this cascaded reset pulse using the procedure shown in Fig.~\ref{fig_3}(a). To find a suitable $\omega _{\mathrm{rst},f}/2\pi$ to reset the $|f\rangle$, we vary $\omega _{\mathrm{rst},f}/2\pi$ while keeping $\omega _{\mathrm{rst},e}/2\pi$ = 3.585 GHz (which is the optimal e-g reset frequency found in Fig.~\ref{fig_2}(b)). The landscape of $P_{f,f}$ versus $t_{\mathrm{rst},f}$ and $\omega_{\mathrm{rst},f}/2\pi$ is shown in Fig.~\ref{fig_3}(b), indicating the optimal $\omega _{\mathrm{rst},f}/2\pi$ to reset the $|f\rangle$ is around 3.801 GHz (dark dashed line in Fig.~\ref{fig_3}(b)). This $\omega _{\mathrm{rst},f}/2\pi$ corresponds to $\omega _{fe}/2\pi \approx \omega _{\mathrm{rst},f}/2\pi -E_C/h$ = 3.590 GHz, which coincides with the resonance frequency of mode A.

With the optimal f-e and e-g reset frequencies obtained in Fig.~\ref{fig_3}(b) and Fig.~\ref{fig_2}(b), we then measure the dependence of reset error on $t_{\mathrm{rst},f}$ for the case when the qubit is initialized to $|e\rangle$ or $|f\rangle$. 
The result is presented in Fig.~\ref{fig_3}(c). 
We set $t_{\mathrm{rst},e}$ = 242 ns, which is the value determined for high-fidelity $|e\rangle$ reset. 
As a result, the $|e\rangle$ reset error $1-P_{g,e}$ shown in Fig.~\ref{fig_3}(c) already reaches its steady value around the readout floor and shows negligible dependence on the f-e reset time $t_{\mathrm{rst},f}$. 
Next, we focus on the $|f\rangle$ reset error $1-P_{g,f}$. 
Similar to the e-g reset process shown in Fig.~\ref{fig_2}, the $1-P_{g,f}$ manifests oscillating decay over $t_{\mathrm{rst},f}$, meaning the f-e reset process is also in the underdamped regime. 
Here, $1-P_{g,f}$ reaches its first minimal (7.3\%) at $t_{\mathrm{rst},f}$ = 64 ns (the vertical dash line in Fig.~\ref{fig_3}(c)). 
This means a f-e-g cascaded reset gate unconditionally resetting both $|e\rangle$ and $|f\rangle$ can be achieved in (64+242) ns = 306 ns. 
The $1-P_{g,f}$ data are fitted using Eq.~(\ref{eq:reset}) (the fit curve is depicted as the red dashed line in Fig.~\ref{fig_3}(c)), yielding $\kappa _{f}^{\prime}/2\pi$ = 9.1 MHz, $g _{qf}^{\prime}/2\pi$ = 5.6 MHz, and the steady-state residual excitation $P_{\mathrm{exc},f}^{\mathrm{s}.\mathrm{s}}$ = 7.1\%. 
The corresponding readout assignment probability matrix (see Appendix~\ref{sec_readout}) shows $1-F_0=7.5\%$, which is comparable to $P_{\mathrm{exc},f}^{\mathrm{s}.\mathrm{s}}$.
This indicates that the measured residual excitation $P_{\mathrm{exc},f}^{\mathrm{s}.\mathrm{s}}$ is mainly limited by the qutrit readout error, rather than reflecting intrinsic excitation remaining after the reset.
Compared to the fitted parameters $\kappa _f/2\pi$ and $g_{qf}/2\pi$ obtained from the e-g reset experiment (see Fig.~\ref{fig_2}), the corresponding values extracted from the f–e reset yield $\kappa _{f}^{\prime}/\kappa _f\approx 1$ and $g_{qf}^{\prime}/g_{qf}\approx \sqrt{2}$, in good agreement with the theoretical prediction \cite{kochChargeinsensitiveQubitDesign2007}. 
The consistency between these two fitting results demonstrates the reliability of our experimental results. 
Note that the filter linewidth $\kappa_f$($\kappa _{f}^{\prime}$) extracted from the time-domain reset dynamics exhibits a slight deviation from that obtained via fitting the S21 curve (see the $\kappa_A$ in Table~\ref{table1}).
The results obtained from the S21 curve fitting may exhibit certain distortions since it is difficult to consider the effects of all the components in the readout chain.

In addition to the cascaded f-e-g reset scheme described above, the relatively small  linewidth of mode A (compared to the transmon anharmonicity) allows us to reset the $|f\rangle$ without damping the $|e\rangle$, which provides a means for implementing the LRU operation that selectively resets the second excited state. 
Here, the LRU operation can be derived from the f-e-g cascaded reset pulses shown in Fig.~\ref{fig_3}(a), by setting $t_{\mathrm{rst},e}$ = 0 ns and only keeping the f-e reset pulse. 
We sweep the $t_{\mathrm{rst},f}$ and measure the transmon state population after the LRU operation. 
As shown in Fig.~\ref{fig_3}(d), the LRU operation can be completed in 62 ns with a residual $|f\rangle$ population of 6.1\%. This relatively large residual population is mainly due to the qutrit readout error.
In Appendix~\ref{sec_LRU} we consider the situations when the LRU operation is applied to $|g\rangle$ and $|e\rangle$, and demonstrate that it has a negligible impact on the coherence within the computational subspace.

Compared to other reset schemes that rely on complex pulse sequences and require precise calibration, the above reset protocol shows advantages in simplicity and stability. 
The reset error remains within an acceptable range even when the reset frequency $\omega_{\mathrm{rst},e}/2\pi$ and $\omega_{\mathrm{rst},f}/2\pi$ slightly deviate from their optimal values (for the numerical evaluation, see Appendix~\ref{sec_paraDrift}). 
This robustness against parameter drift makes it attractive for use in large-scale, multi-qubit processors with automotive control systems.

A potential concern with this reset scheme is that its relatively low reset frequency (3.6 GHz) could enhance the thermal population of the qubit, thereby limiting the achievable reset fidelity.
This limit can be estimated using the Maxwell–Boltzmann distribution
\cite{jinThermalResidualExcitedState2015}
\begin{equation}
p_e/p_g=\exp \left( -\frac{\hbar \omega _{eg}}{k_BT_{\mathrm{eff}}} \right),
\label{eq:thermal}
\end{equation}
where, $p_e$($p_g$) is the excited-state(ground-state) population of the qubit, $\omega_{eg}/2\pi$ is the e-g transition frequency of the qubit, $\hbar$ is the reduced Planck constant and $k_B$ is the Boltzmann constant. 
For $\omega_{eg}/2\pi=3.6\mathrm{GHz}$ and $T_{\mathrm{eff}}=20\mathrm{mK}$, the expected thermal excitation is $p_e=0.02\%$;
at $T_{\mathrm{eff}}=35\mathrm{mK}$, $p_e=0.71\%$.
Since the lowest plate of the dilution refrigerator usually operates below 20 mK, the thermal photon population does not impose a significant limitation on the reset fidelity, provided the device is well thermalized to the refrigerator.

Although we present a detailed characterization of the reset protocol only for a single qubit (Q4) on the filter, the same protocol is expected to reset the other qubits on the filter as well. To illustrate this, Appendix~\ref{sec:otherQrst} provides additional data on the readout and reset characterization of another qubit (Q1) using the same Purcell filter.
For Q1, both the coupling strength $g_{qf}$ and the filter linewidth $\kappa_{f}$ during the reset of the $\ket{e}$ and the leakage reduction unit lie in the underdamped regime.
We note that the effective qubit–filter coupling strength can be tuned toward the critically damped regime via parametric modulation of the qubit frequency \cite{strand2013firstorder,zhouRapidUnconditionalParametric2021}, as shown in Appendix~\ref{sec:Tuning-coupling-strength}.

To evaluate the multiplexing capability of our architecture, we implement a simultaneous reset protocol for qubits Q1 and Q4 coupled to a common Purcell filter (see Appendix~\ref{sec:Simultaneously-resetting-multipl} for details).
To avoid the formation of non-dissipative dark states \cite{marr2003entangledstate} during the reset process, a small frequency detuning is introduced between Q1 and Q4 to break the symmetry protection while keeping both qubits within the linewidth of mode~A.
To support larger detuning ranges and further improve the speed and scalability of the simultaneous reset operation, the linewidth of the fundamental mode should be increased in future designs.
\rev{We note, however, that crosstalk may pose a challenge for this architecture. 
Because multiple qubits are coupled to common filter modes, parasitic interactions, such as residual ZZ interaction, must be carefully addressed before scaling to larger numbers of multiplexed qubits.}

\section{intrinsic Purcell protection}
\label{sec:pfpro}

The qubit in our device has direct capacitive coupling with the filter and with its own readout resonator simultaneously. 
Although the qubit’s idle frequency is designed to be significantly detuned from both the filter modes and the resonator mode, the dissipation induced by this multi-mode electromagnetic environment may still jeopardize the lifetime of the qubit \cite{houckControllingSpontaneousEmission2008}. 
To mitigate this effect, an auxiliary mode (Mode C) is introduced to provide additional Purcell protection near the qubit’s operating frequency, as described in Sec.~\ref{sec:concept} and illustrated in Fig.~\ref{fig_1}.
The effect of this intrinsic Purcell protection will be discussed more quantitatively in this section.

\begin{figure}[ht!]
\centering
\includegraphics[width=0.91\columnwidth]{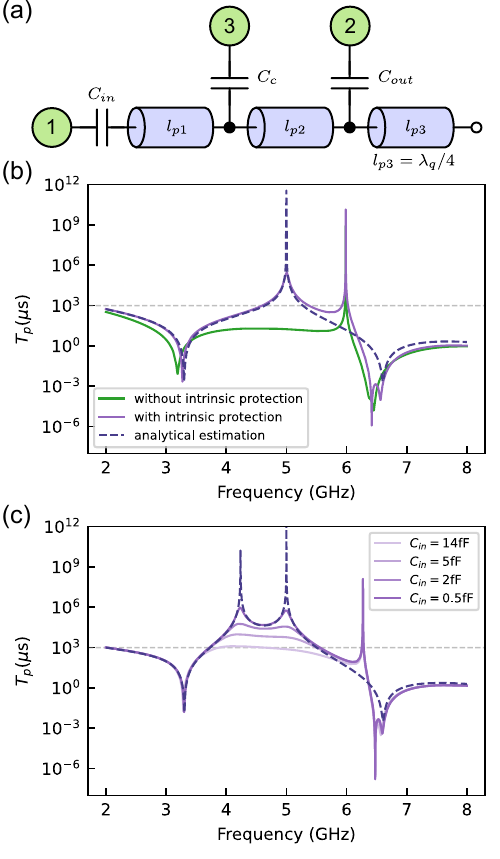}
\caption{
\rev{
Intrinsic Purcell protection of the filter. 
(a) Circuit model of the Purcell filter with additional notch filtering produced by the quarter-wavelength stub. 
Port 1 and Port 2 represent the input and output ports of the filter, respectively. 
Port 3 represents the qubit coupled to the filter. 
(b) The simulated Purcell-limited relaxation times $T_p$ in the cases with and without the intrinsic Purcell filtering. 
The numerical results (solid lines) are obtained from LTspice circuit simulations (Appendix~\ref{sec_simulTp}), while the analytical estimates (dashed lines) follow the model in Appendix~\ref{sec_notchfilterEq}.
The parameters used in the simulation are $C_{in}=14$~fF, $C_{out}=106$~fF, $C_{qf}=2.6$~fF, $C_{fr}=10.1$~fF, $C_{qr}=15$~fF, and $l_{p1}/(l_{p1}+l_{p2}+l_{p3}) \approx 0.01$.
(c) The simulated Purcell-limited relaxation times $T_p$ for different input capacitances $C_{in}$ in the configuration with $l_{p1}/\left( l_{p1}+l_{p2}+l_{p3} \right) \approx 0.39$. 
The simulation method is the same as in (b).
Other parameters used in the simulation are given by $C_{out}=106$~fF, $C_{qf}=2.6$~fF, $C_{fr}=4.4$~fF, and $C_{qr}=15$~fF.
As in (b), the purple dashed curve shows the analytical estimate based on the model in Appendix~\ref{sec_notchfilterEq}.
The simulations in (b) and (c) do not include internal losses of the resonators.
}
}
\label{fig_pfpro}
\end{figure}

We model the filter resonator using the distributed circuit shown in Fig.~\ref{fig_pfpro}, where the filter resonator is separated into three sections with lengths $l_{p1}$, $l_{p2}$, and $l_{p3}$.
Port 1 and port 2 represent the input and the output ports of the filter, respectively, while port 3 represents the qubit coupled to the filter. 
The intrinsic Purcell protection mentioned above can be understood as the blockade of signals transmitted between ports 2 and 3. 
To elucidate this point, we compute the transfer impedance $Z_{23}(\omega)$ between port 2 and port 3, and the solution is given by (see Appendix~\ref{sec_notchfilterEq} for details)
\begin{equation}
Z_{23}=\frac{-jZ_0\cos \beta l_{p1}\cos \beta l_{p3}}{\sin \left( \beta l_p \right)},
\label{eq:z23}
\end{equation}
where $\beta =2\pi /\lambda =\omega /v$ is the propagation constant, $Z_0$ is the characteristic impedance of the transmission line used as the filter resonator, and $l_p=l_{p1}+l_{p2}+l_{p3}$ is the total length of the filter resonator. 
This result shows that at frequencies where $\cos \beta l_{p1}=0$ or $\cos \beta l_{p3}=0$, the transfer impedance between port 2 and port 3 is zero, which means the excitation of the qubit is decoupled with the output port of the filter. 
Here, $l_{p3}$ is designed to be $\lambda_q/4$. 
As a result, the leakage channel between port 2 and port 3 is blocked at the qubit frequency, suppressing the Purcell decay of the qubit. 
If $l_{p3}$ is set to 0, Eq.~(\ref{eq:z23}) degenerates back to the case where there is no intrinsic Purcell protection.

The $T_1$ bound limited by the Purcell effect can be calculated as \cite{esteveEffectArbitraryDissipative1986,houckControllingSpontaneousEmission2008}
\begin{equation}
T_p=C_q/\mathrm{Re}\left[ Y\left( \omega _q \right) \right] ,
\label{eq:tp}
\end{equation}
where $C_q$ is the qubit capacitance, and $Y(\omega_q)$ is the admittance of the external electromagnetic environment as seen by the qubit.
The role of the Purcell protection can be interpreted as a modification of the $Y(\omega_q)$. 
We simulate the Purcell-limited relaxation times $T_p$ for the case with and without the intrinsic Purcell protection using Eq.~(\ref{eq:tp}) (see Appendix~\ref{sec_simulTp} for details), and the results are depicted in Fig.~\ref{fig_pfpro}(b) and (c). 
Analytical predictions derived from the model in Appendix~\ref{sec_notchfilterEq} are plotted for comparison.
We first focus on the result shown in Fig.~\ref{fig_pfpro}(b). 
The numerical simulation reveals a pronounced $T_p$ peak at approximately 5 GHz, which is well captured by the analytical model. 
This $T_p$ peak yields a Purcell-limited relaxation time $T_p$ exceeding 1 ms over a bandwidth of nearly 800 MHz (from 4.59 GHz to 5.37 GHz).
This is a great enhancement compared to the situation without intrinsic Purcell protection, where the $T_p$ is all below 30~$\mu$s from 4.5 GHz to 5.5 GHz. 
Note that apart from the peak around 5 GHz generated by the $\lambda_q/4$ stub, $T_p$ also exhibits an additional peak at approximately 6 GHz (see Fig.~\ref{fig_pfpro}(b)). 
This $T_p$ peak is formed by the interference between the $C_{qr}-C_{fr}$ path and the $C_{qf}$ path \cite{bronnReducingSpontaneousEmission2015}, which can also be used to construct additional Purcell protection.
Since the readout resonator and the $C_{qr}-C_{fr}$ coupling path are not included in the simplified model shown in Appendix~\ref{sec_notchfilterEq}, this additional $T_p$ peak is absent in the analytical prediction.

Apart from the $T_p$ peak at frequency where $\cos \beta l_{p3}=0$, another $T_p$ peak also exists at frequency where $\cos \beta l_{p1}=0$, as shown in Eq.~(\ref{eq:tp}).
In the simulation depicted in Fig.~\ref{fig_pfpro}(b), we consider the case where $l_{p1}/\left( l_{p1}+l_{p2}+l_{p3} \right) \approx 0.01$, i.e., the coupling point is near the input port of the filter. 
Thus, the $T_p$ peak corresponding to $\cos \beta l_{p1}=0$ is located in a relatively high frequency range and cannot be seen in Fig.~\ref{fig_pfpro}(b).
To explore the $T_p$ peak corresponding to $\cos \beta l_{p1}=0$, we carry out another simulation where $l_{p1}/\left( l_{p1}+l_{p2}+l_{p3} \right) \approx 0.39$, and the results are depicted in Fig.~\ref{fig_pfpro}(c).
We can see that apart from the peaks that have been mentioned in Fig.~\ref{fig_pfpro}(b), another $T_p$ peak appears around 4.2 GHz, corresponding to the frequency where $\cos \beta l_{p1}=0$.
The analytical prediction also captures this feature.
\rev{
However, as illustrated in Fig.~\ref{fig_pfpro}(c), $T_{p}$ in this regime is sensitive to the input capacitance $C_{in}$ of the filter, since the mode formed by the first transmission-line section (with length $l_{p1}$) is directly damped by the external microwave circuit through  $C_{in}$.
The analytical model in Appendix~\ref{sec_notchfilterEq} does not consider the damping caused by $C_{in}$.
As a result, the analytical estimation (the purple dashed line) shown in Fig.~\ref{fig_pfpro}(c) corresponds to the situation where $C_{in}\rightarrow 0$.
According to the results shown in Fig.~\ref{fig_pfpro}(c), by choosing a relatively small $C_{in}$, we can achieve a Purcell-limited relaxation time $T_p$ exceeding 1 ms over a bandwidth of about 2 GHz, through the combined intrinsic Purcell protection arising from both $\cos \beta l_{p1}=0$ and $\cos \beta l_{p3}=0$.
}
But we need to keep in mind that the Purcell protection provided by $\cos \beta l_{p1}=0$ depends on the position of the qubit-filter coupling point.
As a result, a general protection provided by $\cos \beta l_{p3}=0$ is still necessary for constructing a Purcell filter shared by multiple qubits.

\begin{figure}[ht!]
\centering
\includegraphics[width=0.91\columnwidth]{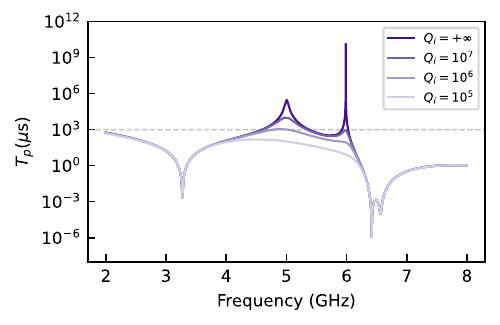}
\caption{
\rev{
Simulated Purcell-limited relaxation times $T_p$ for different internal quality factors $Q_i$. 
Internal loss is incorporated by assigning attenuation to the transmission lines.
Other simulation parameters are identical to those used in Fig.~\ref{fig_pfpro}(b).
}
}
\label{fig_DiffQi}
\end{figure}

\rev{
So far, the internal loss of the resonators has not been taken into account.
To assess its impact, we introduce an attenuation constant $\alpha$ into the transmission lines forming the filter resonators.
The corresponding internal quality factor can then be estimated as $Q_i=\beta/(2\alpha)$~\cite{pozarMicrowaveEngineering}.
The simulation results are shown in Fig.~\ref{fig_DiffQi}.
All circuit parameters are identical to those used in Fig.~\ref{fig_pfpro}(b), except for the inclusion of internal loss.
For $Q_i=10^6$, we obtain $T_p>500\mu$s over a frequency range from approximately 4.5 to 5.25~GHz.
While for $Q_i=10^5$, we only obtain $T_p>150\mu$s about 4.37 to 4.66~GHz.
These results indicate that the internal quality factors of the filter modes and the readout resonators also play an important role in determining the $T_1$ limit of the qubit.
Provided the internal quality factor of the aluminum CPW resonator can reach the level of $10^6$ at single-photon power \cite{megrantPlanarSuperconductingResonators2012}, the multimode Purcell filter proposed here is sufficient for building qubits with $T_1$ above 100~$\mu$s, and is compatible with the long-lived transmon qubits realized recently \cite{placeNewMaterialPlatform2021,wangPracticalQuantumComputers2022,tuokkolaMethodsAchieveNearmillisecond2024,balSystematicImprovementsTransmon2024a,bland2DTransmonsLifetimes2025}.
}

\rev{
In summary, the simulations presented above provide a semi-quantitative explanation for why the qubit energy relaxation times $T_1$ (see Table~\ref{tab:params}) are not significantly degraded when the qubits are coupled to the multimode filter.
Specifically, Fig.~\ref{fig_pfpro}(b) elucidates the operating principle of the intrinsic Purcell protection, while Figs.~\ref{fig_pfpro}(c) and \ref{fig_DiffQi} examine how the input capacitance of the filter, the qubit–filter coupling position, and the internal quality factors of the resonators affect the Purcell-limited $T_1$.
In addition, the impact of the output capacitance $C_\mathrm{out}$ is investigated in Fig.~\ref{fig_DiffCout} (see Appendix~\ref{sec_simulTp} for details).
The circuit parameters used in the simulations do not exactly match those of the fabricated device, but are chosen to be representative of the experimentally accessible parameter range and to clearly illustrate the underlying physical mechanisms.
A full experimental characterization of the Purcell-limited relaxation time $T_\mathrm{p}$ could be achieved using techniques such as driving the qubit through the readout port~\cite{sunadaFastReadoutReset2022,springFastMultiplexedSuperconductingQubit2025}, which we leave for future work.
}
Apart from the energy relaxation $T_1$, we also evaluate the qubit dephasing induced by mode A in Appendix~\ref{sec_dephasing}, and find that this influence is negligible for our device parameters since the detuning between mode A and the qubit is much larger than their coupling strength.

\section{Discussion}
\label{sec:discussion}

Our results demonstrate that distinct qubit operations, such as fast unconditional reset and dispersive readout, can be realized using different-order modes of a single microwave resonator, thereby eliminating the need for additional on-chip components.
This hardware-efficient approach, implemented via a multi-mode Purcell filter, not only preserves qubit coherence through intrinsic Purcell protection, but also provides fast dissipative channels essential for active reset and leakage removal, which are useful for quantum error correction.
Although higher-order modes in coplanar waveguide resonators are often considered as parasitic modes and detrimental to coherence \cite{houckControllingSpontaneousEmission2008}, our work shows that they can be systematically engineered into functional resources. 
This design principle—assigning distinct qubit functionalities to different-order modes of a single multi-mode resonator—opens a new direction for scaling up superconducting quantum processors, particularly in architectures constrained by wiring and footprint.

Using this multi-mode filter, we demonstrate rapid reset of the $|e\rangle$ and $|f\rangle$ states within 220 ns and 310 ns.
In addition, an LRU operation that selectively resets the second excited state can be completed in 62 ns. 
Other reset schemes using parametric modulations \cite{zhouRapidUnconditionalParametric2021,kimFastUnconditionalReset2024} or adiabatic swap operations \cite{mcewenRemovingLeakageinducedCorrelated2021} are also feasible on our device. 
In the experiment described in Sec.~\ref{sec:reset}, the reset process is in the underdamped regime where $g_{qf}>\kappa _f/4$. 
To further improve the reset speed, the linewidth of the filter can be increased to reach the critically damped regime $g_{qf}=\kappa _f/4$. 
The characterization of the reset error is mainly limited by the readout fidelity, which can be improved by optimizing the dispersive shift and the effective linewidth of the resonator to reach larger state separation \cite{walterRapidHighFidelitySingleShot2017,blaisCircuitQuantumElectrodynamics2021b}, introducing quantum-limited amplifiers \cite{cavesQuantumLimitsNoise1982,castellanos-beltranAmplificationSqueezingQuantum2008,eichlerQuantumLimitedAmplificationEntanglement2014,mutusStrongEnvironmentalCoupling2014,macklinQuantumLimitedJosephson2015,kaufmanJosephsonParametricAmplifier2023a}, engineering nonperturbative cross-Kerr interactions \cite{yeUltrafastSuperconductingQubit2024,wang999fidelityMeasuringSuperconducting2024,chappleBalancedCrossKerrCoupling2025}, using a shelving scheme that exploits higher energy levels of the qubit \cite{elderHighFidelityMeasurementQubits2020,malletSingleshotQubitReadout2009,chenTransmonQubitReadout2023,wang999fidelityMeasuringSuperconducting2024,jiangGeneration95qubitGenuine2025,zhangDemonstratingQuantumError2025}, and implementing more sophisticated readout pulses \cite{mcclureRapidDrivenReset2016,chenTransmonQubitReadout2023,jergerDispersiveQubitReadout2024,swiadekEnhancingDispersiveReadout2024}. 
Our multi-mode Purcell filter can be extended to a multi-stage configuration \cite{yanBroadbandBandpassPurcell2023a}. 
In this case, the filter can have a large passband for accommodating readout resonators, which is attractive for use in large-scale quantum processors.


Note that the Purcell effect in the multi-mode environment should be carefully considered, and additional Purcell protection may be necessary as described in Sec.~\ref{sec:pfpro}. 
At the same time, the multi-mode environment provided by the filter may also serve as a useful platform for exploring multi-mode circuit quantum electrodynamics. 
Furthermore, the lossy fundamental mode of the filter in our device can be used as a resource for research on dissipative systems \cite{hanExceptionalEntanglementPhenomena2023}.

\section*{Acknowledgements}

This work was supported by the
National Natural Science Foundation of China (Grants
No. 92265207, No. T2121001,  No. T2322030, No. 12122504, No. 12274142,
No. 92365206, No. 12104055, No. 12475017, No. U25A6009), the Natural Science Foundation of Guangdong Province (Grant No. 2024A1515010398),
the Innovation Program for Quantum Science and
Technology (Grants No. 2021ZD0301800,
No. 2021ZD0301802), the Beijing Nova Program
(No. 20220484121). We thank the support from the Synergetic Extreme Condition User Facility (SECUF) in Huairou District, Beijing. Devices were made at the Nanofabrication Facilities at Institute of Physics, CAS in Beijing.

\section*{Data availability}
The data are available from the authors upon reasonable request.

\appendix	
\section{Device information and experimental setup}
\label{sec_device}

\begin{table}[htb]
\caption{\label{tab:params}
Qubit and readout parameters. Here, $\omega_{eg}$ and $\alpha$ are the transmon e-g transition frequency and anharmonicity, respectively. $T_1$ is the qubit lifetime around its idle point. The readout resonator parameters include the frequency $\omega_{r}$, full dispersive shift $2\chi$, and linewidth $\kappa_{r}$.
}
\begin{ruledtabular}
\begin{tabular}{lccccc} 
 & Q1 & Q2 & Q4 & Q5 & Q6 \\
\hline
$\omega_{eg}/2\pi$ (GHz) & 4.706 & 4.619 & 4.457 & 4.578 & 4.574 \\ 
$\alpha/2\pi$ (MHz) & -207 & - & -211 & -211 & -209 \\
$T_1$ ($\mathrm{\mu s}$) & 42 & 59 & 58 & 75 & 52 \\
$\omega_{r}/2\pi$ (GHz) & 6.421 & 6.442 & 6.486 & 6.517 & 6.530 \\
$2\chi/2\pi$ (MHz) & 3.0 & 2.8 & 2.4 & 1.7 & 1.1 \\
$\kappa_r/2\pi$ (MHz) & 1.2 & 0.4 & 0.9 & 0.4 & 0.4 \\
\end{tabular}
\end{ruledtabular}
\end{table}

\begin{figure}[ht!]
\centering
\includegraphics[width=0.91\columnwidth]{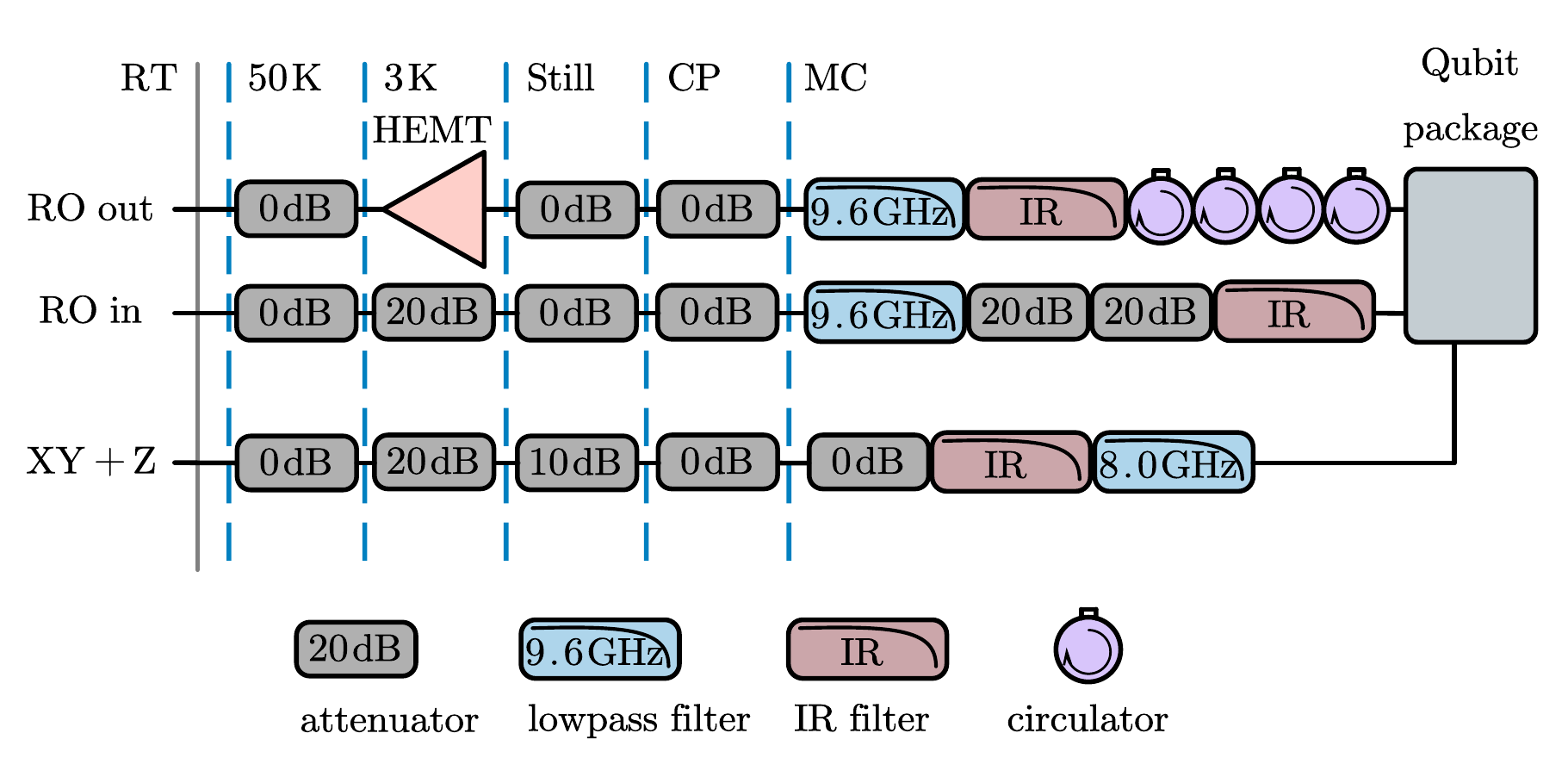}
\caption{ 
Schematic of the wiring setup. The XY and Z control lines are integrated into one line. The readout pulse is sent in through the readout-input line (labelled as RO in), and the scattered signal containing state information leaves the chip through the readout-output line (labelled as RO out).
}
\label{fig_wiring}
\end{figure}

Our experiment is carried out on a flip-chip superconducting processor, consisting of lithographically patterned aluminum films on sapphire substrates. 
The transmon qubits are fabricated on the top chip, and the readout resonators, the multi-mode Purcell filter, and the control lines are integrated on the carrier chip. 
The details of device fabrication can be found in Ref.~\cite{liuPrethermalizationRandomMultipolar2025}. 
The multi-mode Purcell filter is shared by six qubits and their corresponding readout resonators.
Their parameters are summarized in Table~\ref{tab:params}, in which the full dispersive shift $2\chi$ and the effective linewidth $\kappa_r$ of the readout resonators are extracted by the 'chi-kappa power' (CKP) experiment  \cite{sankSystemCharacterizationDispersive2025}.
For clarity, we label the qubits according to their physical positions along the filter: Q1 is closest to the input port, Q6 is closest to the output port, and Q2-Q5 are arranged sequentially in between.
Q3 is not included in Table~\ref{tab:params} because it was likely damaged during fabrication.
The device is cooled down in a dilution refrigerator, and the wiring setup is shown in Fig.~\ref{fig_wiring}. 
Note that no quantum-limited amplifier is induced in the readout chain.

\section{Extracting the parameters of the multimode filter}
\label{sec_filterFit}

\begin{figure}[ht!]
\centering
\includegraphics[width=0.91\columnwidth]{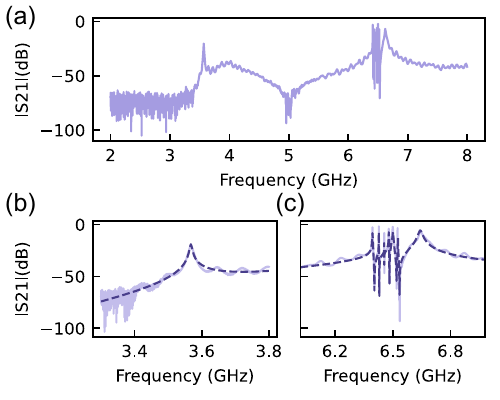}
\caption{ 
(a) Transmission spectrum S21 of the readout chain collected by a vector network analyzer.
(b) S21 of the frequency range covering the fundamental mode (Mode A) of the filter. 
(c) S21 of the frequency range covering the second-order mode (Mode B) of the filter.
The dashed lines in (b) and (c) are the fits using Eq.~(\ref{eq:s21}).
}
\label{fig_s21fit}
\end{figure}

The transmission coefficient of a bandpass filter capacitively coupled to the readout line can be calculated following the input-output theory \cite{naamanSynthesisParametricallyCoupled2022,seteQuantumTheoryBandpass2015b,heinsooRapidHighfidelityMultiplexed2018} and the results can be written as (here we assumed $n$ readout resonators are coupled to the filter)
\begin{equation}
S_{21}=\frac{-j\kappa _f/2}{\left( j\Delta _{fd}+\kappa _f/2 \right) +\sum_{i=1}^n{|g_{fr,i}|^2/\left( j\Delta _{r_id}+\gamma _{r,i}/2 \right)}},
\label{eq:s21}
\end{equation}
where $\kappa_f/2\pi$ is the linewidth of the filter,  $\Delta _{fd}=\omega _f-\omega _d$ ($\Delta _{r_id}=\omega _{r_i}-\omega _d$) is the detuning between the filter(the readout resonators) and the probe tune, $g_{fr, i}$ is the coupling strength between the filter and the readout resonators, and $\gamma_{r, i}$ is induced to account for the intrinsic loss of the $i-\mathrm{th}$ readout resonators. 
Although this formula just involves a single mode of the filter, it still well describes the transmission spectrum of our multimode filter, as long as the fundamental mode and the second-order mode of the filter are treated separately. 
To do that, the transmission data around the resonance frequency of the fundamental mode and the second-order mode are extracted and fitted to Eq.~(\ref{eq:s21}) individually. 
To account for the influence of components like the amplifiers on the readout chain, a linear background is added to the fit function: $20\log _{10}|\mathrm{S}_{21}|+kf_{\mathrm{d}}+b$. 
The fit curves are shown in Fig.~\ref{fig_s21fit}(b) and (c), and the extracted parameters are listed in Table~\ref{table1}.

\section{Resonance frequency shift induced by shunt stubs}
\label{sec:f_shift}

\begin{figure}[ht!]
\centering
\includegraphics[width=0.91\columnwidth]{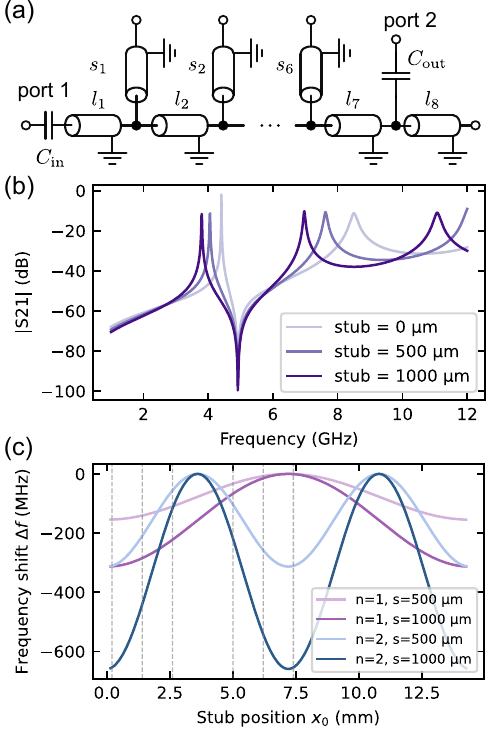}
\caption{
Impact of $C_{qf}$ stubs on the filter resonance frequency.
(a) A filter resonator interrupted by six shunt stubs. The filter is formed by transmission line segments labeled $l_1, l_2, \dots, l_8$, while the shunt stubs are labeled $s_1, s_2, \dots, s_6$.
(b) Transmission spectrum $S_{21}$ of the two-port network in (a) for different stub lengths. The lengths of $l_1$–$l_8$ are set to 200, 1200, 1200, 2400, 1200, 1200, 500, and 6500~$\mu\mathrm{m}$, respectively. The six shunt stubs are identical in length. The relative dielectric constant is chosen to be $\epsilon_{eff}=5.5$. 
(c) Analytical estimation of the resonance frequency shift $\Delta f$ of the fundamental ($n=1$) and second-order ($n=2$) modes induced by a shunt stub, calculated by Eq.~(\ref{eq:delta_f}). 
The total filter length is 14400~$\mu\mathrm{m}$, consistent with (b), and the stub length is labeled as $s$. The vertical dashed lines indicate the positions of the $C_{qf}$ stubs in our device.
}
\label{fig_f_shift}
\end{figure}

As discussed in the main text, the branches used to connect the $C_{qf}$ (hereafter referred to as $C_{qf}$ stubs) modify the resonance frequency of the filter resonator, and different-order modes of the filter experience different frequency shifts. This explains why the frequency of mode B is not exactly twice that of mode A. 
In this appendix, we provide a quantitative analysis of this effect.

To this end, we model the filter architecture using the two-port network shown in Fig.~\ref{fig_f_shift}(a), where a transmission-line resonator is interrupted by six open-circuited stubs representing the $C_{qf}$ stubs.
The transmission spectrum $S_{21}$ of this network is obtained by cascading the ABCD matrices of the individual transmission-line sections ($l_1, l_2, …, l_7$), the shunt stubs ($s_1, s_2, …, s_6$), the final line $l_8$, and the input/output coupling capacitors.
The calculated spectra are shown in Fig.~\ref{fig_f_shift}(b). 
Increasing the length of the $C_{qf}$ stubs lowers the resonance frequencies, and the shift of the second-order mode is larger than that of the fundamental mode.
When the stub length reaches approximately 1000 $\mu\mathrm{m}$, the calculated $S_{21}$ roughly coincides with the measured data (Fig.~\ref{fig_1}(d).
Although the physical length of the $C_{qf}$ stub on our device is less than 1000 $\mu\mathrm{m}$, it is still a reasonable value accounting for the effective length provided by the indium bump connector.
From Fig.~\ref{fig_f_shift}(b), we also note that even with zero stub length, the frequency of mode B is still not exactly double that of mode A. 
That is because the stub $l_8$ used to construct intrinsic Purcell protection also has a similar effect on filter resonance frequency.

To understand why shunt stubs induce different frequency shifts for mode A and mode B, we consider a simplified scenario: a filter resonator of length $L$ with a single open-circuited stub of length $s$ placed at position $x_0$. 
The admittance of the shunt stub with length $s$ can be expressed as $jY_s=j\tan \left( \beta s \right) /Z_0$. 
Applying Kirchhoff’s Current Law at position $x_0$ (where $V_{x_0}$ is the voltage at this position) gives 
\begin{equation}
\frac{V_{x_0}}{-jZ_0\cot \left( \beta x_0 \right)}+\frac{V_{x_0}}{-jZ_0\cot \left( \beta \left( L-x_0 \right) \right)}+jY_sV_{x_0}=0,
\end{equation}
i.e.,
\begin{equation}
\tan \left( \beta x_0 \right) +\tan \left( \beta \left( L-x_0 \right) \right) =-Y_sZ_0.
\label{eq:stub}
\end{equation}
We define the left-hand side of the Eq.~(\ref{eq:stub}) as $F\left( \beta \right) =\tan \left( \beta x_0 \right) +\tan \left( \beta \left( L-x_0 \right) \right) $, and the first-order Taylor expansion of $F(\beta)$ around the resonance point $\beta_0L=n\pi$ yields
\begin{equation}
F\left( \beta _0+\Delta \beta \right) \simeq F\left( \beta _0 \right) +\Delta \beta F^\prime\left( \beta _0 \right) =\Delta \beta F^\prime\left( \beta _0 \right) .
\end{equation}
Substituting this into the Eq.~(\ref{eq:stub}) gives
\begin{align}
\Delta \beta &\simeq -\frac{Y_s\!\left(\beta _0\right) Z_0}{F'\!\left(\beta _0\right)} \notag \\
&= -\frac{Y_s\!\left(\beta _0\right) Z_0}{x_0\sec ^2\!\left(\beta _0 x_0\right)
+\left(L-x_0\right)\sec ^2\!\left[\beta _0\left(L-x_0\right)\right]} ,
\end{align}
from which we finally obtain the resonance frequency shift caused by the shunt stub:
\begin{align}
\Delta f & = \frac{v}{2\pi}\Delta \beta \notag \\
&= -\frac{v}{2\pi}\frac{Y_s\!\left(\beta _0 \right) Z_0}{x_0\sec ^2\!\left(\beta _0x_0 \right)
+\left( L-x_0 \right) \sec ^2\!\left[\beta _0\left( L-x_0 \right) \right]} \notag \\
&= -\frac{v}{2\pi}\frac{\tan \!\left( \beta _0s \right) Z_0}{x_0\sec ^2\!\left(\beta _0x_0 \right)
+\left( L-x_0 \right) \sec ^2\!\left[\beta _0\left( L-x_0 \right) \right]} ,
\label{eq:delta_f}
\end{align}
where $\beta_0L=n\pi$ and $v=c/\sqrt{\epsilon _{eff}}$.
Using this equation, we calculate the resonance frequency shift $\Delta f$ as a function of the stub position $x_0$ and length $s$, with the results presented in Fig.~\ref{fig_f_shift}(c).
This figure highlights two key trends: (i) for a given mode, the shift is largest when the stub is located at a voltage antinode and smallest at a voltage node; and (ii) for the same stub configuration, higher-order modes generally experience larger frequency shifts than lower-order modes.
The positions of the $C_{qf}$ stubs in our device, indicated by vertical dashed lines, mostly result in a larger $\Delta f$ for the second-order mode than for the fundamental mode. 
This explains the observation in Fig.~\ref{fig_f_shift}(b), where mode B is more sensitive to the $C_{qf}$ stubs than mode A, leading to the fact that mode B’s frequency is not exactly twice that of mode A.

\section{Readout characterization and error analysis}
\label{sec_readout}

\begin{figure}[ht!]
\centering
\includegraphics[width=0.91\columnwidth]{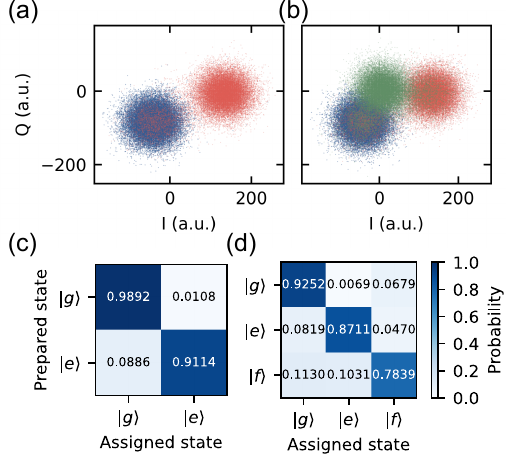}
\caption{ 
Readout characterization of Q4. (a) and (b) are the IQ signals in the cases of the two-level state readout and the three-level state readout, respectively. The $|g\rangle$, $|e\rangle$, and $|f\rangle$ are represented by blue dots, red dots, and green dots, respectively. 30000 repetitions of the experiment are performed for each prepared state. The corresponding assignment probability matrices are presented in (c) and (d).
}
\label{fig_readout}
\end{figure}

The readout assignment probability matrices of the qubit used in the reset experiment are presented in Fig.~\ref{fig_readout}. 
The $|g\rangle$ and $|e\rangle$ assignment errors in the case of two-level state discrimination are $\varepsilon _0$ = 1.08\% and $\varepsilon _1$ = 8.86\%, respectively. 
To analyze the source of the readout errors, we further divide the errors into two categories: the separation error and the state error \cite{sankFastAccurateState2014}. 
The separation error reflects the intrinsic signal-to-noise ratio of the dispersed photons and can be captured by the overlap area of the Gaussian fits to the signal clusters in the IQ plane. 
Here, the separation errors of the two-level state discrimination are found to be $\varepsilon _{s,0}$ = 0.38\% and $\varepsilon _{s,1}$ = 0.08\%. 
The remaining part $\varepsilon _{t,0\left( 1 \right)}=\varepsilon _{0\left( 1 \right)}-\varepsilon _{s,0\left( 1 \right)}$ is the state error caused by state-transition events before or during the readout process. 
The finite $T_1$ of the qubit is a common source of the $|e\rangle$ state error $\varepsilon _{t,1}$. 
In our experiment, the average $T_1$ of the qubit is around 50~$\mu$s.
For a readout integration time $\tau_m$ = 2 $\mu$s, the bound of $\varepsilon _1$ can be estimated as
\begin{eqnarray}
\varepsilon _{T1}=&&1-\exp \left( -\tau _m/T_1 \right) 
\nonumber\\
\approx&& 1-\left( 1-\tau _m/T_1 \right) =\tau _m/T_1=4\%.
\end{eqnarray}
The remaining part of the state error $\varepsilon _{t,1}-\varepsilon _{T1}$ is mainly due to the measurement-induced state transition. 
We suspect that dephasing noise at the qubit-readout detuning frequency \cite{boissonneaultDispersiveRegimeCircuit2009,slichterMeasurementInducedQubitState2012} may be responsible for the measurement-induced state transition observed during the readout process.
The $|g\rangle$ state error $\varepsilon _{t,0}=\varepsilon _0-\varepsilon _{s,0}$ = 0.70\% may be caused by the thermal excitation.
Compared to the two-level state readout, the three-level state readout has relatively larger assignment errors, which is partly due to the relatively large overlap of these three clusters of states in the IQ plane, as shown in Fig.~\ref{fig_readout}(b).

\section{Reset characterization of an additional qubit on the filter}
\label{sec:otherQrst}

\begin{figure}[ht!]
\centering
\includegraphics[width=0.91\columnwidth]{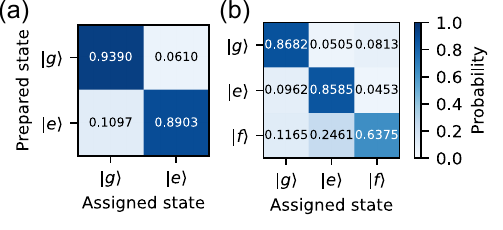}
\caption{ 
The readout assignment matrices of Q1. 15000 repetitions of the experiment are performed for each prepared state.
}
\label{fig_Q1readout}
\end{figure}

\begin{figure}[ht!]
\centering
\includegraphics[width=0.91\columnwidth]{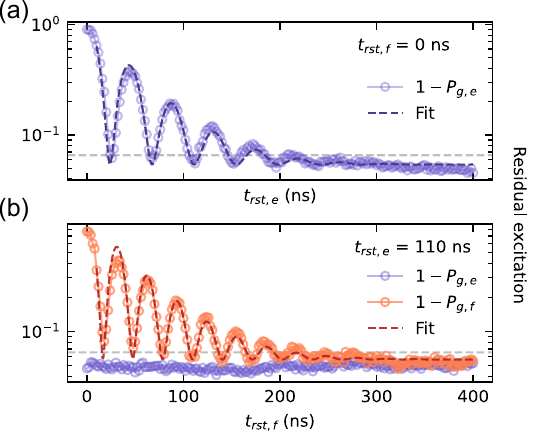}
\caption{
Residual excitation of Q1 as a function of the reset time.
(a) and (b) correspond to the e-g and f-e-g reset processes, respectively.
The readout floor $1-F_0$ is indicated by the horizontal dashed lines.
}
\label{fig_Q1rst}
\end{figure}

\begin{table}[htb]
\caption{\label{tab:rst}
Linewidth and coupling strength extracted from the reset data.
$\kappa_f$($\kappa _{f}^{\prime}$) and $g_{qf}$($g _{qf}^{\prime}$) denote the parameters obtained by fitting the e-g (f-e) reset data.
}
\begin{ruledtabular}
\begin{tabular}{lcccc} 
 & $\kappa_f/2\pi$(MHz) & $g_{qf}/2\pi$(MHz) & $\kappa _{f}^{\prime}/2\pi$(MHz) & $g _{qf}^{\prime}/2\pi$(MHz) \\
\hline
Q1 & 7.2 & 11.6 & 7.0 & 16.2 \\ 
Q4 & 8.5 & 3.9 & 9.1 & 5.6 \\
\end{tabular}
\end{ruledtabular}
\end{table}

To provide an additional example beyond the Q4 results shown in Sec.~\ref{sec:reset}, we present in this appendix the reset characterization of Q1, as displayed in Fig.~\ref{fig_Q1rst}, which complements the main results and indicates that reset can be extended to other qubits sharing the same filter.

Both the e-g and f-e reset processes of Q1 exhibit underdamped dynamics, where the population decays with oscillations.
For the e-g reset shown in Fig.~\ref{fig_Q1rst}(a), the third minimum provides a suitable operating point that balances reset time and fidelity, yielding a residual excitation of $1-P_{g,e}=5.9\%$ in 110 ns.
For the f-e-g reset in Fig.~\ref{fig_Q1rst}(b), the third minimum again offers an appropriate choice, corresponding to $1-P_{g,f}=6.3\%$ with $t_{\mathrm{rst},f}=80\mathrm{ns}$.
Taken together, this enables unconditional reset of both $|e\rangle$ and $|f\rangle$ within 190~ns.

Following the procedure in Sec.~\ref{sec:reset}, we fit the e-g and f-e reset data using Eq.~\ref{eq:reset}, and the extracted parameters are summarized in Table~\ref{tab:rst}.
Compared with Q4, Q1 is located closer to the voltage antinode of mode A, resulting in a stronger coupling to this mode.

\section{Simultaneously resetting multiple qubits}\label{sec:Simultaneously-resetting-multipl}

\begin{figure}
\includegraphics[scale=0.7]{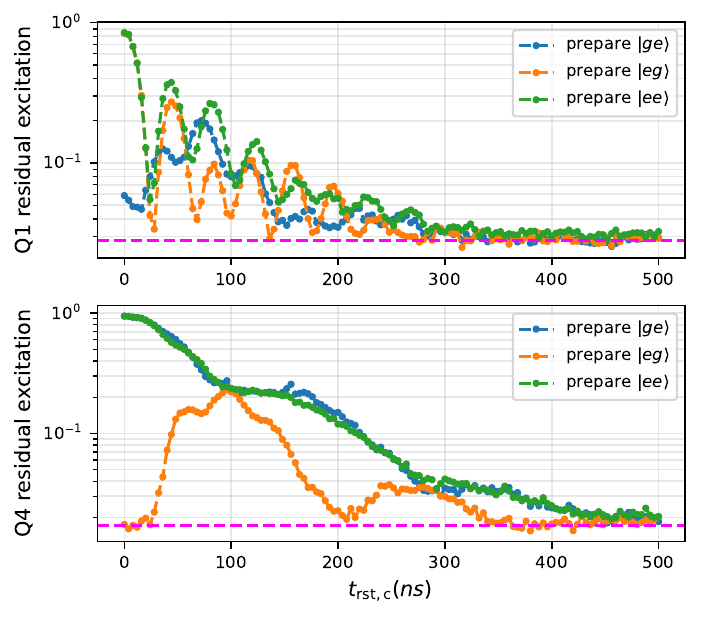} 
\caption{Residual excitation of resetting both Q1 and Q4 simultaneously with
the qubits initially prepared in $|ge\rangle$, $|eg\rangle$ and
$|ee\rangle$. The dashed magenta lines indicate the measured probability
when the qubits are fully reset for a long enough duration. }\label{fig:simul-reset}
\end{figure}

We experimentally demonstrate simultaneous reset of two qubits, Q1 and Q4, coupled to a common Purcell filter. 
In Fig. \ref{fig:simul-reset}, the qubits are tuned to be near-resonant with the fundamental mode, and hold for a common duration $t_{{\rm rst,c}}$. 
To circumvent the formation of non-dissipative dark states that typically emerge when
qubits are on resonance \cite{marr2003entangledstate}, we introduce
a small detuning between Q1 and Q4 to break the symmetry protection while maintaining their frequencies within the linewidth of mode A during reset. 
The reset frequencies of Q1 and Q4 are set to $3.527$ GHz and $3.572$ GHz, respectively. 
We prepare the initial states $|ge\rangle$, $|eg\rangle$ and $|ee\rangle$ before performing the reset operation. 
It is observed that the excitation of one qubit can be transferred to the other during reset operation.
Although this inter-qubit exchange could undermine the reset fidelity for the qubit initially in $|g\rangle$, all of the residual excitation for the prepared states are reset to $\le1\%$ above the convergent probability after 400 ns.

The relatively small qubit detuning used in this experiment is constrained by the narrow linewidth of mode~A, which is designed to perform leakage reduction units without disturbing the computational subspace.
This narrow linewidth, however, limits the speed of the simultaneous reset. 
In our future designs, the reset speed can be improved by increasing the linewidth of the fundamental mode,
albeit at the expense of reduced selectivity in resetting the $\ket{f}$ state without perturbing the computational subspace. 
Nevertheless, in such cases, leakage reduction units on data qubits could still be realized through alternative strategies, such as transferring leakage from data qubits to auxiliary qubits \cite{miaoOvercomingLeakageQuantum2023}.

\section{Tuning coupling strength with parametric modulation}\label{sec:Tuning-coupling-strength}

\begin{figure*}
\begin{centering}
\includegraphics[scale=0.7]{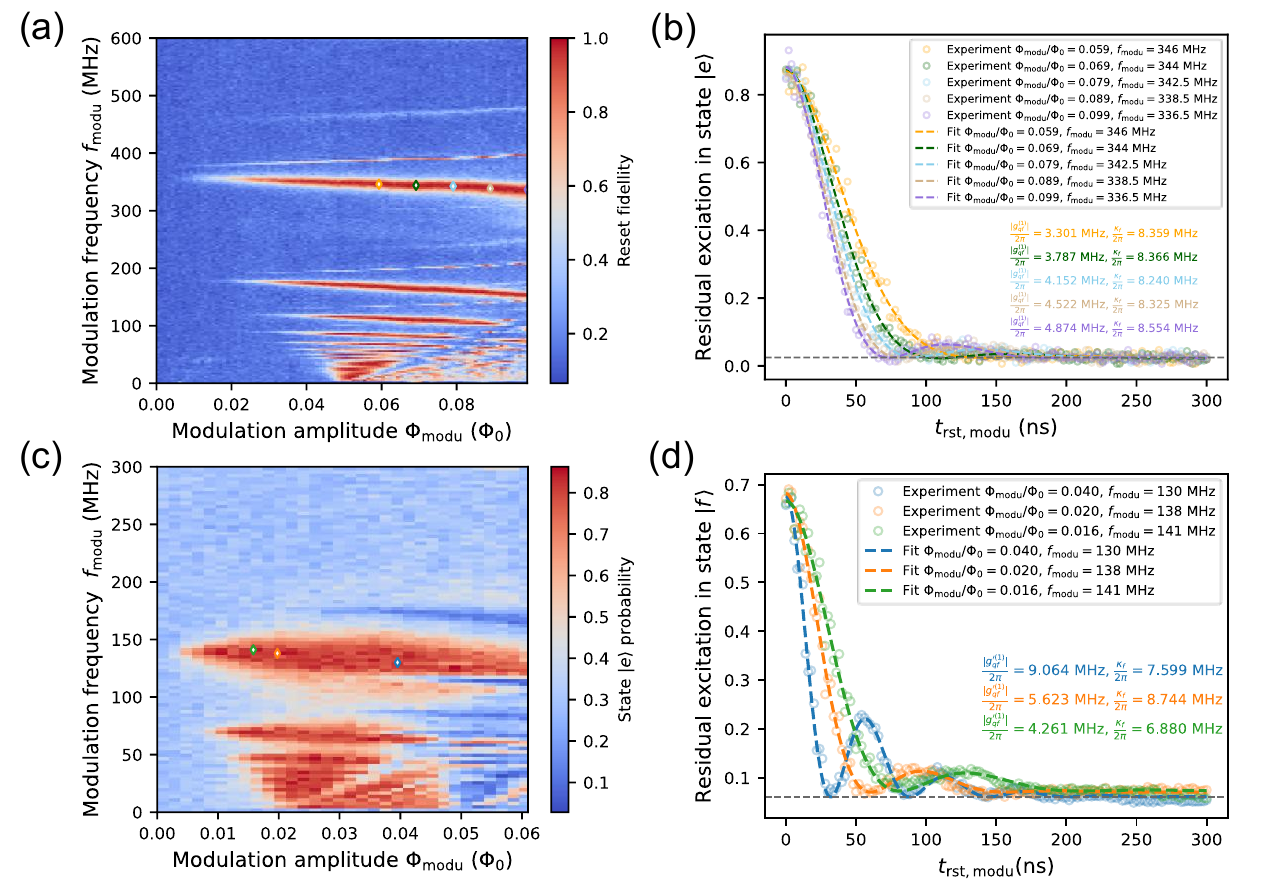}
\par\end{centering}
\caption{Parametric reset and leakage reduction operations of Q1. 
(a) Reset fidelity of the qubit initially prepared in $\ket{e}$ as a function of modulation amplitude and frequency.
(b) Residual excitation versus reset time for the parameters indicated in (a).
(c) Probability of the $\ket{e}$ during parametric leakage reduction for different modulation amplitudes and frequencies.
(d) Probability of the $\ket{f}$ as a function of time under parametric leakage reduction.
The horizontal dashed gray lines indicate the readout floor. 
}
\label{fig:Parametric-reset-of-q1}
\end{figure*}

For the qubit-filter swap in the underdamped regime, as is the case for Q1, the first several minima of reset error may be above threshold required. 
Fortunately, our reset schemes can be combined with parametric modulation to tune the effective coupling rate toward the critically damped regime ($g_{qf} = \kappa_f / 4$), thereby enhancing the robustness of the reset process.

In our implementation of the parametric reset protocol, the qubit is first prepared in the excited state $\ket{e}$, and its average transition frequency is tuned to $\overline{f_{eg}} = 3.915~\mathrm{GHz}$. A sinusoidal modulation is then applied to the qubit frequency, described by
$f_{eg}(t) = \overline{f_{eg}} + A \cos(2\pi f_{\mathrm{modu}} t)$,
where the modulation amplitude $A$ is determined by the magnetic flux $\Phi_{\mathrm{modu}}$ biasing the qubit.
To identify the optimal reset parameters, we first measure the reset fidelity as a function of modulation amplitude and frequency.
As shown in Fig.\ref{fig:Parametric-reset-of-q1}(a), multiple sidebands can be brought into resonance with the filter’s fundamental mode at different modulation frequencies.
To enable the maximum coupling strength, we choose the first-order sideband to perform reset, with an effective coupling strength $g_{qf}^{(1)}$ given by
$|g_{qf}^{(1)}| = |g_{qf} J_1(A / f_{\mathrm{modu}})|$,
where $J_1$ is the first-order Bessel function \cite{zhouRapidUnconditionalParametric2021,strand2013firstorder}.
The parameters indicated by the markers in Fig.~\ref{fig:Parametric-reset-of-q1}(a)
are selected for scanning the residual excitation over different reset times
[Fig.~\ref{fig:Parametric-reset-of-q1}(b)], from which $|g_{qf}^{(1)}|$ and $\kappa_f$ are extracted by fitting the decay curves. 
It can be inferred from the decay characteristics and parameters fitting results that the effective coupler strength $|g_{qf}^{(1)}|$ demonstrates
a monotonic increase with the modulation amplitude. 
The results show that the effective coupling strength increases monotonically with the modulation amplitude, since the argument $A / f_{\mathrm{modu}}$ has not reached the first maximum of $J_1$ in the positive domain.

Parametric modulation can also be applied to leakage reduction units, as demonstrated in Fig\ref{fig:Parametric-reset-of-q1}(c).
The parameters marked in the figure are used to measure the $\ket{f}$-state dynamics shown in Fig.~\ref{fig:Parametric-reset-of-q1}(d). Different effective coupling strengths
$|g_{qf}^{\prime(1)}| = \sqrt{2} |g_{qf}^{(1)}|$
for the $\ket{f_q 0_f} \leftrightarrow \ket{e_q 1_f}$ exchange are obtained by adjusting the modulation amplitude and corresponding frequency.

The above parameteric reset scheme based on the filter's fundamental mode provides advantages over conventional schemes using far-detuned readout resonators \cite{zhouRapidUnconditionalParametric2021}. 
Since we can bias the qubit frequency closer to the filter's fundamental mode, parametrically modulating
qubit frequency at this operating point requires narrower sidebands,
which avoids potential unwanted interaction between qubits that could arise from different orders of sidebands during reset operations.

\section{Numerical evaluation of robustness to reset frequency drift}
\label{sec_paraDrift}

In Sec.~\ref{sec:reset}, we demonstrate the e-g and f-e-g reset scheme by applying Z pulse to tune the qubit frequency. 
In this appendix, we further examine the robustness of these schemes against inaccuracies in the reset frequencies. 
With numerical simulation, we show that reset error remains within an acceptable range even when the reset frequencies deviate slightly from their optimal values.

To simulate the reset process in Sec.~\ref{sec:reset}, we introduce a circuit QED model written as
\begin{align}
&H = H_q + H_f + H_{int}, \\
&H_q/\hbar = \omega_{eg}(t) b^{\dagger}b + \frac{\alpha}{2} b^{\dagger} b^{\dagger} b b, \\
&H_f/\hbar = \omega_f a^{\dagger} a, \\
&H_{int}/\hbar = g_{qf} \left( a^{\dagger} b + a b^{\dagger} \right),
\end{align}
where $a^{\left( \dagger \right)}$ and $b^{\left( \dagger \right)}$ are the creation and annihilation operators for the filter mode and the qubit. 
In this model, the transmon qubit is treated as a nonlinear oscillator with a resonance frequency $\omega _{eg}\left( t \right)$ and an anharmonicity $\alpha$, and the filter mode used to reset the qubit is described as a linear oscillator with a resonance frequency $\omega_f$ and a coupling strength of $g_{qf}$ to the qubit. 
For simplicity, we have not included the readout resonator in the model, since the readout resonator is far detuned with both the qubit and the filter mode (the fundamental mode) during the reset process. 
The effect of coupling to the environmental degrees of freedom can be described by the master equation
\begin{equation}
\dot{\rho} = -\frac{i}{\hbar}\left[ H,\rho \right] 
+ \kappa_f \mathcal{D}[a] \rho 
+ \gamma_q \mathcal{D}[b] \rho ,
\label{eq:master_equ}
\end{equation}
where $\kappa_f$ is the linewidth of the filter mode, $\gamma_q$ is the qubit decay rate, and $\mathcal{D} \left[ L \right] \rho =\left( 2L\rho L^{\dagger}-L^{\dagger}L\rho -\rho L^{\dagger}L \right) /2$. 
To verify the reliability of this model, we use Eq.~(\ref{eq:master_equ}) to simulate the reset dynamics, with the parameters extracted from the fit of the e-g reset process in Sec.~\ref{sec:reset} and the qubit operator $b^{\left( \dagger \right)}$ truncated to three levels. 
To faithfully describe the experimental characterization limited by finite readout error, we also introduce an steady-state reset error of 0.8\% and 7.1\% for the e-g and the f-e-g reset processes, respectively. 
The results of the simulation show great agreement with the experimental data, as shown in Fig.~\ref{fig_2}(d) and Fig.~\ref{fig_3}(c).

\begin{figure}[ht!]
\centering
\includegraphics[width=0.91\columnwidth]{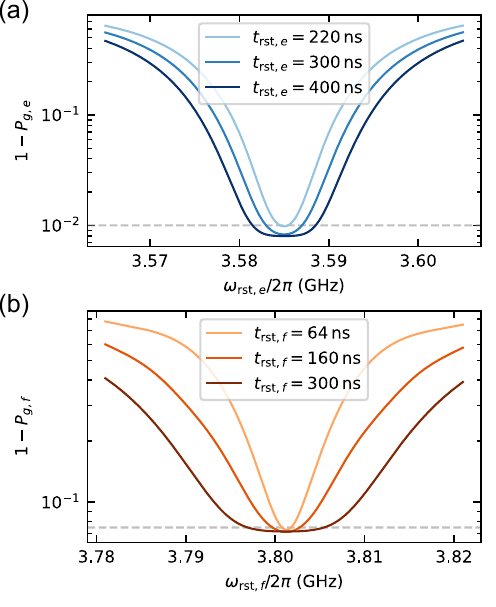}
\caption{ 
Numerical evaluation of the impact of reset-frequency inaccuracy.
(a) Reset error $1-P_{g,e}$ as a function of the reset frequency $\omega _{\mathrm{rst},e}/2\pi$, obtained from simulations based on Eq.~(\ref{eq:master_equ}) using the fitted parameters of the e–g reset process in Sec.~\ref{sec:reset}. 
A residual error of 0.8\% is included to account for factors like readout infidelity. 
The horizontal dashed line indicates a reset error of 1\%.
(b) Reset error $1-P_{g,f}$ of the f–e–g reset process versus the reset frequency $\omega _{\mathrm{rst},f}/2\pi$, simulated with the same method and parameters as in (a). 
The e-g reset process uses $\omega _{\mathrm{rst},e}/2\pi=3.585$ GHz and  $t_{\mathrm{rst},e}=240$~ns. 
A residual error of 7.1\% is added in the simulation. 
The horizontal dashed line corresponds to the readout floor $1-F_0$ = 7.5\%.
}
\label{fig_w_rst}
\end{figure}

To evaluate the impact of flux-bias inaccuracy, we simulate the e–g reset scheme with the reset frequency scanned around its optimal value of 3.585 GHz. 
The resulting residual excitation is presented in Fig.~\ref{fig_w_rst}(a). 
The reset error remains below 2\% when the frequency deviates from its optimal value by nearly $\pm3$ MHz, with a reset time of $t_{\mathrm{rst},e}=220$~ns. 
When the reset time is extended to 400 ns, the error can be maintained below 1\% over a wider frequency range, from 3.582 GHz to 3.588 GHz. 
A similar analysis is performed for the f–e–g reset process, with the results shown in Fig.~\ref{fig_w_rst}(b). 
In this case, the reset error approaches the readout error floor when the reset frequency deviates from its optimum by approximately $\pm1$ MHz and $\pm5$ MHz, for reset times of $t_{\mathrm{rst},f}=64$~ns and $t_{\mathrm{rst},f}=300$~ns, respectively. 
These results indicate that our reset protocol exhibits considerable tolerance to inaccuracies in the reset frequency. 
At the same time, this robustness depends on the chosen reset time: longer reset durations enhance robustness but may limit overall operation speed. 
Therefore, an appropriate reset time should be selected according to the specific requirements of the experiment.

\section{Impact of the LRU on the computational subspace}
\label{sec_LRU}

\begin{figure}[ht!]
\centering
\includegraphics[width=0.91\columnwidth]{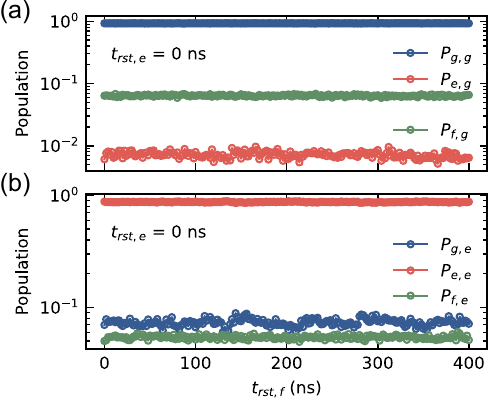}
\caption{ 
The state populations of a transmon during the LRU operation, where $t_{\mathrm{rst},e}=0$, $\omega _{\mathrm{rst},f}/2\pi=3.801$ GHz, and $t_{\mathrm{rst},f}$ is swept. 
The pulse parameters are identical to those used in Fig.~\ref{fig_3}(d).
Panels (a) and (b) correspond to initial states $|g\rangle$ and $|e\rangle$, respectively. 
}
\label{fig_lru_ge}
\end{figure}

A LRU is defined \cite{aliferisFaultTolerantQuantumComputation2006,battistelHardwareEfficientLeakageReductionScheme2021} as an operation such that (1) the leakage faults can be reduced to regular faults within the computational subspace after the application of the LRU, (2) the coherence within the computational subspace is negligibly affected when the LRU is applied to a nonleaked state.
In Sec.~\ref{sec:reset} we demonstrate that our LRU operation can deplete the population in $|f\rangle$ [Fig.~\ref{fig_3}(d)]. 
Here in this section, we show that it has a negligible impact on the coherence within the computational subspace.

Fig.~\ref{fig_lru_ge} displays the effect of the LRU on a transmon initialized in $|g\rangle$ and $|e\rangle$, respectively. 
\rev{The data shown in Fig.~\ref{fig_lru_ge} are measured under the same conditions as those in Fig.~\ref{fig_3}(d), except for a different initial state.}
We can see that although some small fluctuations in $P_{e,g}$ and $P_{g,e}$ are observed, the state populations remain essentially unaffected, and no significant reduction of lifetime is detected when the transmon is initialized in $|e\rangle$.
These results confirm that the LRU operation negligibly affects states within the computational subspace.

\section{The intrinsic notch filtering produced by the quarter-wavelength stub}
\label{sec_notchfilterEq}

\begin{figure}[ht!]
\centering
\includegraphics[width=0.91\columnwidth]{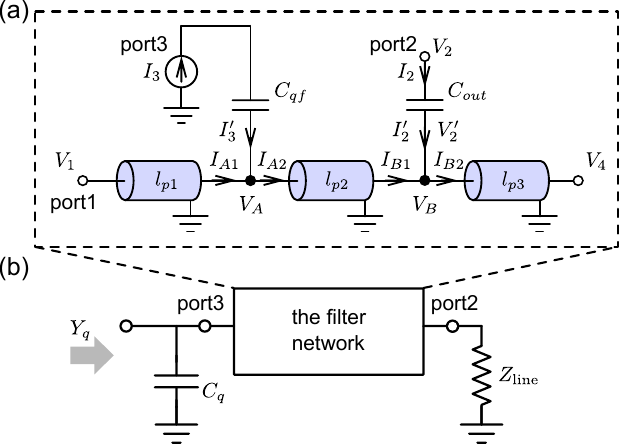}
\caption{ 
(a) Distributed-circuit model of the filter with a quarter-wavelength stub. The Purcell filter is divided into three segments with lengths $l_{p1}$, $l_{p2}$ and $l_{p3}$, respectively. 
The input and output ports of the filter are labeled as port 1 and port 2, respectively. 
Port 3 represents the qubit coupled to the filter.
(b) Circuit model for deriving the $T_p$ of the qubit. The filter network is the one presented in (a). Port 3 of the network is connected to a qubit with shunt capacitance $C_q$, and port 2 of the network is connected to the readout line with impedance 
$Z_{\mathrm{line}}$. The input admittance of the circuit seen by the qubit is labeled as $Y_q$.
}
\label{fig_z23}
\end{figure}

As described in Sec.~\ref{sec:concept} and Sec.~\ref{sec:pfpro}, by moving the output capacitor to the position $\lambda_q/4$ away from the open end of the filter, a quarter-wavelength stub is constructed, which prevents the microwave signal whose frequency is around $\omega_q/2\pi$ dissipating into the readout line and provides the additional Purcell protection. 
This working principle can also be formulated by computing the transfer impedance $Z_{23}=\frac{V_2}{I_3}\mid_{I_k=0\ \mathrm{for}\ k\ne3}$ between the coupling point and the output port of the filter \cite{springFastMultiplexedSuperconductingQubit2025}. 
Here we will calculate the transfer impedance $Z_{23}$ of the model in Fig.~\ref{fig_z23}(a) following the theory in \cite{pozarMicrowaveEngineering}. And the analytical prediction of the Purcell-limited relaxation time $T_p$ is also formulated in this appendix.

Since the transmission(ABCD) matrix of a series capacitance $C$ is
\begin{equation}
T=\left( \begin{matrix}
	1&		Z\\
	0&		1\\
\end{matrix} \right) =\left( \begin{matrix}
	1&		\frac{1}{j\omega C}\\
	0&		1\\
\end{matrix} \right) ,
\end{equation}
the voltages and currents at nodes A and B in Fig.~\ref{fig_z23}(a) are given by
\begin{equation}
I_{3}^{\prime}=I_3,
\end{equation}
\begin{equation}
0=I_2=I_{2}^{\prime},
\end{equation}
\begin{equation}
V_2
=V_{2}^{\prime}+I_{2}^{\prime}/\left( j\omega C_{out} \right) =V_{2}^{\prime}=V_B.
\end{equation}
The input impedance of a lossless transmission line terminated in an open-circuited end can be written as
\begin{eqnarray}
Z_{in}\left( z \right) =&&\frac{V\left( z \right)}{I\left( z \right)}=Z_0\frac{Z_L+jZ_0\tan \beta z}{Z_0+jZ_L\tan \beta z}\mid_{Z_L=+\infty}^{}\nonumber\\
=&&Z_0\frac{1}{j\tan \beta z},
\end{eqnarray}
by which we get the input impedance of point A(B) looking towards the open-circuited end 1(4) in Fig.~\ref{fig_z23}(b):
\begin{align}
\frac{V_A}{-I_{A1}} &= Z_0 \frac{1}{j \tan(\beta l_{p1})} \equiv Z_A, \\
\frac{V_B}{ I_{B2}} &= Z_0 \frac{1}{j \tan(\beta l_{p3})} \equiv Z_B.
\end{align}
Using Kirchhoff’s Current Law at nodes A and B gives
\begin{align}
I_{A2} &= I_{A1}+I_{3}^{\prime}, \\
I_{B2} &= I_{B1}+I_{2}^{\prime} = I_{B1}.
\end{align}
The transmission line section $l_{p2}$ between node A and node B can be seen as a two-port network with the ABCD matrix given by
\begin{align}
T_{lp2} 
&= \begin{pmatrix}
	A_2 & B_2 \\
	C_2 & D_2
\end{pmatrix} 
\nonumber \\[6pt]
&= \begin{pmatrix}
	\cos \beta l_{p2} & jZ_0 \sin \beta l_{p2} \\
	jY_0 \sin \beta l_{p2} & \cos \beta l_{p2}
\end{pmatrix},
\label{eq:Tlp2}
\end{align}
by which we get
\begin{align}
V_A=A_2V_B+B_2I_{B1}, \\
I_{A2}=C_2V_B+D_2I_{B1}.
\end{align}

Then, combining the above relations, the current at port 3 can be written as
\begin{align}
I_3 &= I_{3}^{\prime} = I_{A2}-I_{A1} = C_2 V_B + D_2 I_{B1} + \frac{V_A}{Z_A} \nonumber \\
&= C_2 V_B + \frac{D_2 V_B}{Z_B} + \frac{A_2 V_B + \tfrac{B_2 V_B}{Z_B}}{Z_A} \nonumber \\
&= \left( C_2 + \frac{D_2}{Z_B} + \frac{A_2}{Z_A} + \frac{B_2}{Z_A Z_B} \right) V_B.
\label{eq:I3}
\end{align}
As a result, the transfer impedance $Z_{23}$ can be calculated as
\begin{align}
Z_{23} &= \frac{V_2}{I_3} 
= \frac{V_B}{I_3} \nonumber \\
& = \frac{1}{A_2/Z_A + \tfrac{B_2}{Z_A Z_B} + C_2 + D_2/Z_B} \nonumber \\
&= \frac{-j Z_0 \cos(\beta l_{p1}) \cos(\beta l_{p3})}{\sin(\beta l_p)} ,
\label{eq:Z23}
\end{align}
which is the Eq.~(\ref{eq:z23}) in the main text. Using similar methods, $Z_{22}$ and $Z_{33}$ of the filter network in Fig.~\ref{fig_z23}(a) can also be formulated by
\begin{align}
Z_{22}&=-j\left( Z_0\frac{\cos \left( \beta l_{p3} \right) \cos \left[ \beta \left( l_p-l_{p3} \right) \right]}{\sin \left( \beta l_p \right)}+\frac{1}{\omega C_{out}} \right), \\
Z_{33}&=-j\left( Z_0\frac{\cos \left( \beta l_{p1} \right) \cos \left[ \beta \left( l_p-l_{p1} \right) \right]}{\sin \left( \beta l_p \right)}+\frac{1}{\omega C_{qf}} \right) .
\end{align}

Next, with the impedance matrix elements derived above, the Purcell-limited relaxation time $T_p$ of the qubit can be analytically estimated using the model shown in Fig.~\ref{fig_z23}(b). In this model, the filter is viewed as a two-port network where its input and output ports are connected to the qubit and the readout line, respectively. And its ABCD matrix can be calculated using the impedance matrix elements derived above:
\begin{equation}
T_F=\left( \begin{matrix}
	A_F&		B_F\\
	C_F&		D_F\\
\end{matrix} \right) =\left( \begin{matrix}
	Z_{33}/Z_{23}&		\frac{Z_{33}Z_{22}-Z_{23}^{2}}{Z_{23}}\\
	1/Z_{23}&		Z_{22}/Z_{23}\\
\end{matrix} \right) .
\end{equation}
With this ABCD matrix, the input admittance of the circuit seen by the qubit (labeled as $Y_q$ in Fig.~\ref{fig_z23}(b)) can be formulated as
\begin{equation}
Y_q= j\omega C_q+\frac{C_FZ_{\mathrm{line}}+D_F}{A_FZ_{\mathrm{line}}+B_F} .
\end{equation}
Then $T_p$ of the qubit can be obtained using Eq.~(\ref{eq:tp}), and the results are plotted in Fig.~\ref{fig_pfpro} in the main text.

\section{Simulation of the Purcell-limited relaxation times}
\label{sec_simulTp}

\begin{figure}[ht!]
\centering
\includegraphics[width=0.91\columnwidth]{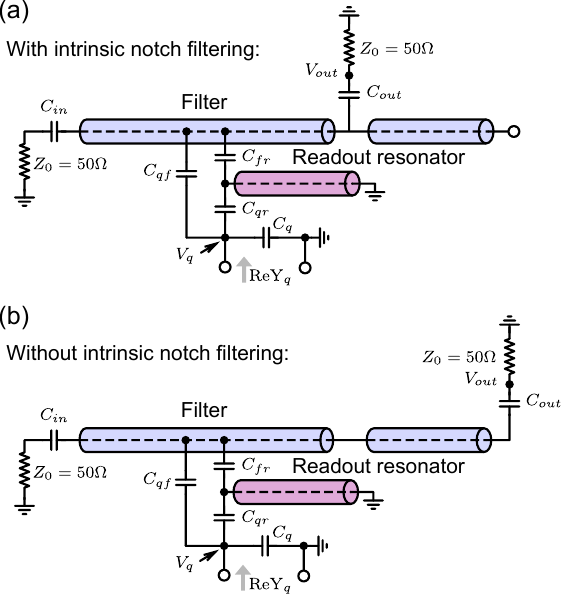}
\caption{ 
Circuit models used for numerical simulation. (a) and (b) are the situations with and without the intrinsic notch filtering produced by the quarter-wavelength stub, respectively.
}
\label{fig_Tp_simul}
\end{figure}

The Purcell decay rate of a transmon qubit coupled to a lossy linear mode off-resonantly is given by $\Gamma =\kappa \left( g/\Delta \right) ^2$ \cite{kochChargeinsensitiveQubitDesign2007}, where $g$ is the coupling strength between the qubit and the linear mode, $\kappa$ is the effective linewidth of the mode, and $\Delta$ is the detuning. 
However, this formula only considers a single EM mode and fails to describe the Purcell decay rate of the qubit here, which is damped in a multi-mode environment provided by our multi-mode filter. 
To estimate the Purcell decay rate of the qubit in our device, we follow an alternative method in which the classical admittance $Y_q$ of the external circuit seen by the qubit is calculated and thus the Purcell-limited relaxation time $T_p$ is given by Eq.~(\ref{eq:tp}) in the main text.
Here, $Y(\omega_q)$ is calculated using numerical simulation of the circuit in LTspice. 
The circuit models are shown in Fig.~\ref{fig_Tp_simul}(a) and Fig.~\ref{fig_Tp_simul}(b), which correspond to the situations with and without the intrinsic Purcell protection, respectively. 
To account for the finite internal quality factor $Q_i$ of the filter resonator, an attenuation constant $\alpha$ is added to the transmission lines used to construct the filter resonator.
The corresponding quality factor is given by $Q_i=\beta/(2\alpha)$ \cite{pozarMicrowaveEngineering}.
In the simulation, the qubit is replaced by a voltage source $V_q$. 
By measuring the corresponding source current $I_q$, the real part of the admittance is simply given by $\mathrm{Re}Y_q=\mathrm{Re}\left( I_q/V_q \right)$. 
Following this way, we calculate the $T_p$ of the qubit with and without the intrinsic Purcell protection, and obtain the results shown in Fig.~\ref{fig_pfpro} and Fig.~\ref{fig_DiffQi} in the main text.

\begin{figure}[ht!]
\centering
\includegraphics[width=0.91\columnwidth]{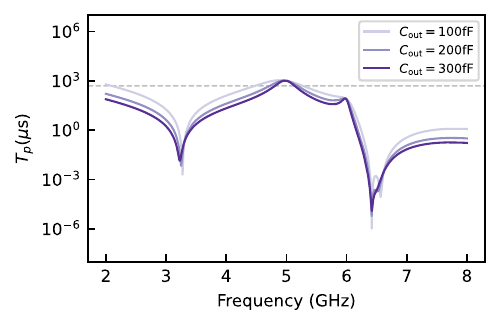}
\caption{ 
\rev{
Simulated Purcell-limited relaxation times $T_p$ for different output capacitances $C_{\mathrm{out}}$ of the filter.
In the simulation the internal quality factors of the filter and readout resonators are assumed to be $Q_i=10^6$ .
Other parameters used in the simulations are consistent with those in Fig.~\ref{fig_pfpro}(b).
The horizontal dashed line corresponds to $T_p=500\mu$s.
}
}
\label{fig_DiffCout}
\end{figure}

\rev{
Here we provide additional simulations of the Purcell-limited relaxation time $T_\mathrm{p}$ for different output capacitances $C_\mathrm{out}$, as shown in Fig.~\ref{fig_DiffCout}.  
In our current device, $C_\mathrm{out}\!\approx\!100$ fF.  
Increasing $C_\mathrm{out}$ to 200 fF and 300 fF broadens the filter linewidth by approximately fourfold and ninefold, respectively, indicating its scalability to larger systems with more readout resonators multiplexed onto a single readout line.
However, as also shown in Fig.~\ref{fig_DiffCout}, increasing $C_\mathrm{out}$ reduces the bandwidth over which strong Purcell protection is maintained.
Specifically, the bandwidth of the frequency region satisfying $T_\mathrm{p} > 500~\mu$s decreases to approximately 450~MHz and 320~MHz when $C_\mathrm{out}$ is increased to 200~fF and 300~fF, respectively.
As a result, the number of readout resonators that can be multiplexed onto a single filter must be carefully balanced against the achievable level and bandwidth of Purcell protection, highlighting a trade-off between scalability and qubit lifetime.
}

\section{Dephasing induced by mode A}
\label{sec_dephasing}

\begin{figure}[ht!]
\centering
\includegraphics[width=0.91\columnwidth]{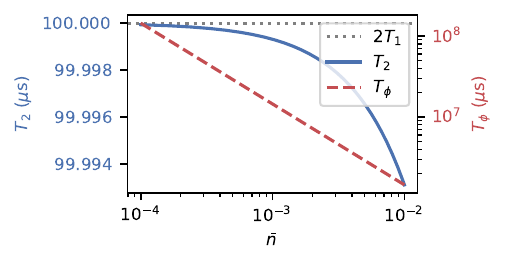}
\caption{ 
Photon-noise-induced dephasing rate due to mode~A, calculated using Eq.~(\ref{eq:dephasing}). 
The corresponding decoherence time limited by mode~A is evaluated as 
$T_{2}=\bigl[(T_{\phi})^{-1}+(2T_{1})^{-1}\bigr]^{-1}$ with $T_{1}=50~\mu\text{s}$.
}
\label{fig_dephas}
\end{figure}

Since the qubit is coupled to the filter directly, low-frequency modes of the filter, such as mode~A, may be thermally populated and cause photon-noise-induced dephasing of the qubit. In the appendix, we estimate the influence of this effect.

For weak thermal noise, the photon-noise-induced dephasing rate is given by \cite{gambettaQubitphotonInteractionsCavity2006,sunadaPhotonNoiseTolerantDispersiveReadout2024}
\begin{equation}
\Gamma _{\phi}=\frac{\kappa \left( 2\chi \right) ^2}{\kappa ^2+\left( 2\chi \right) ^2}\bar{n},
\label{eq:dephasing}
\end{equation}
where $\kappa$ is the linewidth of the resonator, $2\chi$ is the full dispersive shift, and $\bar{n}$ represents the average noise-photon number. 
In the reset experiment described in Sec. \ref{sec:reset}, mode A was measured to have a linewidth of $\kappa/2\pi$ = 8.5 MHz and a coupling strength to the qubit of $g_{qf}/2\pi$ = 3.9 MHz.
The qubit idle frequency and mode~A frequency are $\omega_{eg}/2\pi=4.5$~GHz and $\omega_{A}/2\pi=3.6$~GHz, respectively, yielding a dispersive shift of approximately $2\chi/2\pi=0.01$~MHz. 
Using these parameters, the induced dephasing time $T_{\phi}$ can be calculated from Eq.~\ref{eq:dephasing}, and the result is plotted in Fig.~\ref{fig_dephas}.
As shown in Fig.~\ref{fig_dephas}, the dephasing induced by mode~A remains very small even when the noise-photon number reaches $\bar{n}\sim10^{-2}$ (by comparison, the best reported values are on the order of $\bar{n}\sim10^{-4}$ \cite{sunadaPhotonNoiseTolerantDispersiveReadout2024}).  
This is mainly because the detuning between mode~A and the qubit is much larger than their coupling strength, which strongly suppresses the dispersive shift $2\chi$ to a value far below the linewidth $\kappa$.  
Consequently, the dephasing induced by mode~A is expected to be negligible for our device parameters.


\bibliography{cite}

\end{document}